%% file: NGC1377_accepted.tex
\documentclass{aa}  

\usepackage{graphicx}
\usepackage{txfonts}
\usepackage{natbib}
\usepackage{url}
\usepackage{amsmath}
\usepackage{amsmath}
\usepackage{multirow,bigdelim}
\begin{document}

   \title{Radio continuum and X-ray emission from the most extreme FIR-excess galaxy NGC~1377}

   \subtitle{An extremely obscured AGN revealed}

   \author{F. Costagliola
          \inst{1,2}
          \and
          R. Herrero-Illana\inst{3}
	  \and
	  A. Lohfink\inst{4}
          \and
	  M. P\'erez-Torres\inst{3,5}
	  \and
	  S. Aalto\inst{1}
	  \and
	  S. Muller\inst{1}
	  \and
	  A. Alberdi\inst{3}
          }

   \institute{Chalmers University of Technology, Onsala Space Observatory, SE-439 92 Onsala, Sweden, \email{costagli@chalmers.se}
              \and INAF-Istituto di Radioastronomia - Italian ARC, via Gobetti 101, 40129, Bologna, Italy 
              \and Instituto de Astrof{\'i}sica de Andaluc{\'i}a (IAA-CSIC), Glorieta de la Astronom{\'i}a, s/n, E-18008, Granada, Spain
	      \and Institute of Astronomy, University of Cambridge, Madingley Road, Cambridge CB30HA, UK
	      \and Visiting Scientist: Departamento de F\'isica Teorica, Facultad de Ciencias, Universidad de Zaragoza, Spain
             }

 
  \abstract
   {Galaxies which strongly deviate from the radio-far IR correlation are of great importance for studies of galaxy evolution as they may be tracing early, short-lived stages of starbursts and active galactic nuclei (AGNs). The most extreme FIR-excess galaxy NGC~1377 has long been interpreted as a young dusty starburst, but millimeter observations of CO lines revealed a powerful collimated molecular outflow which cannot be explained by star formation alone.}
   {To determine the nature of the energy source in the nucleus of NGC~1377 and to study the driving mechanism of the collimated CO outflow.}
   {We present new radio observations at 1.5 and 10~GHz obtained with the Jansky Very Large Array (JVLA)  and \textit{Chandra} X-ray observations towards NGC~1377. The observations are compared to synthetic starburst models to constrain the properties of the central energy source.}
   {We obtained the first detection of the cm radio continuum and X-ray emission in NGC~1377. We find that the radio emission is distributed in two components, one on the nucleus and another offset by 4$''$.5 to the South-West. We confirm the extreme FIR-excess of the galaxy, with a $q_\mathrm{FIR}\simeq$4.2, which deviates by more than 7-$\sigma$ from the radio-FIR correlation. Soft X-ray emission is detected on the off-nucleus component. From the radio emission we estimate for a young ($<10$~Myr) starburst a star formation rate SFR$<$0.1~M$_\odot$~yr$^{-1}$. Such a SFR is not sufficient to power the observed IR luminosity and to drive the CO outflow.}
   {We find that a young starburst cannot reproduce all the observed properties of the nucleus of NGC~1377. We suggest that the galaxy may be harboring a radio-quiet, obscured AGN of 10$^6$M$_\odot$, accreting at near-Eddington rates. We speculate that the off-nucleus component may be tracing an hot-spot in the AGN jet.}


   \maketitle
%

\section{Introduction}

\begin{figure}
\centering
\includegraphics[width=.5\textwidth,keepaspectratio]{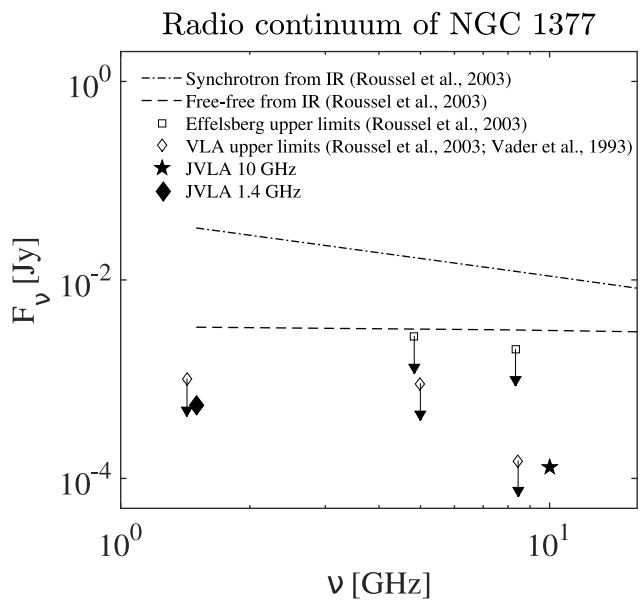}
\caption{\label{fig:sed} Radio continuum flux of NCG1377 compared to free-free and synchrotron emission expected from IR fluxes. Filled symbols refer to our detections of the radio continuum with the JVLA. The upper limits are 3-$\sigma$ estimates from the literature. }
\end{figure}

A small fraction of galaxies \citep[e.g., ][]{helou_85} have faint radio and bright far-infrared (FIR) emission which strongly deviate from the well-known radio to far-infrared (FIR) correlation \citep[e.g., ][]{condon92}. Potential interpretations of the FIR excess include very young synchrotron-deficient starbursts or dust-enshrouded active galactic nuclei (AGN). FIR-excess galaxies only represent a small sub-group ($\approx$1$\%$) of the IRAS Faint Galaxy Sample \citep[e.g., ][]{roussel03}, which is likely an indication that they may be tracing a short evolutionary phase. If powered by obscured AGN, these are likely in the early stages of their evolution, when the nuclear material has not yet been dispersed and/or consumed to feed the growth of the black hole. Recent publications \citep{aalto1377,sakamoto2013,aalto2016} have shown that some of such systems drive molecular outflows and are thus the ideal targets to study the first stages of starburst/AGN feedback.

The most extreme FIR-excess galaxy detected so far is NGC~1377, a lenticular galaxy in the Eridanus group at a distance of 21~Mpc (1$''$ = 102 pc), with a FIR luminosity of the order of 10$^{9}$~L$_\odot$ \citep{roussel03}. Despite several attempts in the past, the radio continuum in this galaxy has long remained undetected. Deep observations with the Very Large Array (VLA) and Effelsberg telescopes \citep[e.g., ][ and references therein]{roussel03} obtained limits on the radio continuum which are about 40 times fainter than the synchrotron emission that would be expected from the radio to far-infrared correlation (Fig.~\ref{fig:sed}). Also, these limits are fainter than the free-free emission that would be expected from the star formation rate of 1-2 M$_\odot$ yr$^{-1}$ derived from the IR flux. HII regions are not detected through near-infrared hydrogen recombination lines or thermal radio continuum \citep{roussel03,roussel06}. Deep mid-infrared silicate absorption features suggest that the nucleus is very compact, and enshrouded by a large mass of dust \citep[e.g., ][]{spoon07}, which potentially absorbs all ionizing photons. The high obscuration makes the determination of the energy source a challenging task. 

\citet{roussel06} proposed that NGC~1377 is a nascent opaque starburst -- the radio synchrotron deficiency would then be caused by the extreme youth (pre-supernova stage) of the starburst activity when the young stars are still embedded in their birth-clouds. In contrast, \citet{imanishi06} argued, based on the small 3.3~$\mu$m PAH equivalent width and strong mid-IR H$_2$ emission, that NGC~1377 harbors a buried AGN. Furthermore, \citet{imanishi09} found an HCN/HCO$^+$ J = 1--0 line ratio exceeding unity, which they suggested is evidence of an X-ray dominated region (XDR) surrounding an AGN. The authors explained the lack of radio continuum with the presence of a large column of intervening material that causes free-free absorption. 

With recent SMA and ALMA observations, \citet{aalto1377,aalto2016} found a molecular outflow originating from the inner 30 pc of NGC~1377 and extending to about 150--200~pc. Given its velocity and extent, these authors calculated an age for the outflow of about 1.4~Myr, which is consistent with the young age of the central activity. The upper limit on the 1.4~GHz flux density falls short by a factor of 10 to explain the outflow as supernova-driven and the authors suggest instead that the outflow may be driven by radiation pressure from a buried AGN.

In summary, there is substantial evidence that the energy source of NGC~1377 must be young, but its nature is still highly debated. Here we report the results of recent observations with the Enhanced Karl G. Jansky Very Large Array (JVLA) and \textit{Chandra} X-ray observatory which finally reveal the energy source of NGC~1377 and shed new light on the properties of FIR-excess galaxies. The details of the observations are reported in Section~\ref{sec:obs}. In Section~\ref{sec:res} we describe the properties of the radio and X-ray emission and compare the observations with synthetic starburst models. In Section~\ref{sec:disc} we discuss the nature of the nuclear energy source and in Section~\ref{sec:conc} we summarize our conclusions.
   
\section{Observations and data reduction}
\label{sec:obs}
\begin{figure*}
\begin{center}
\includegraphics[height=.5\textwidth,keepaspectratio]{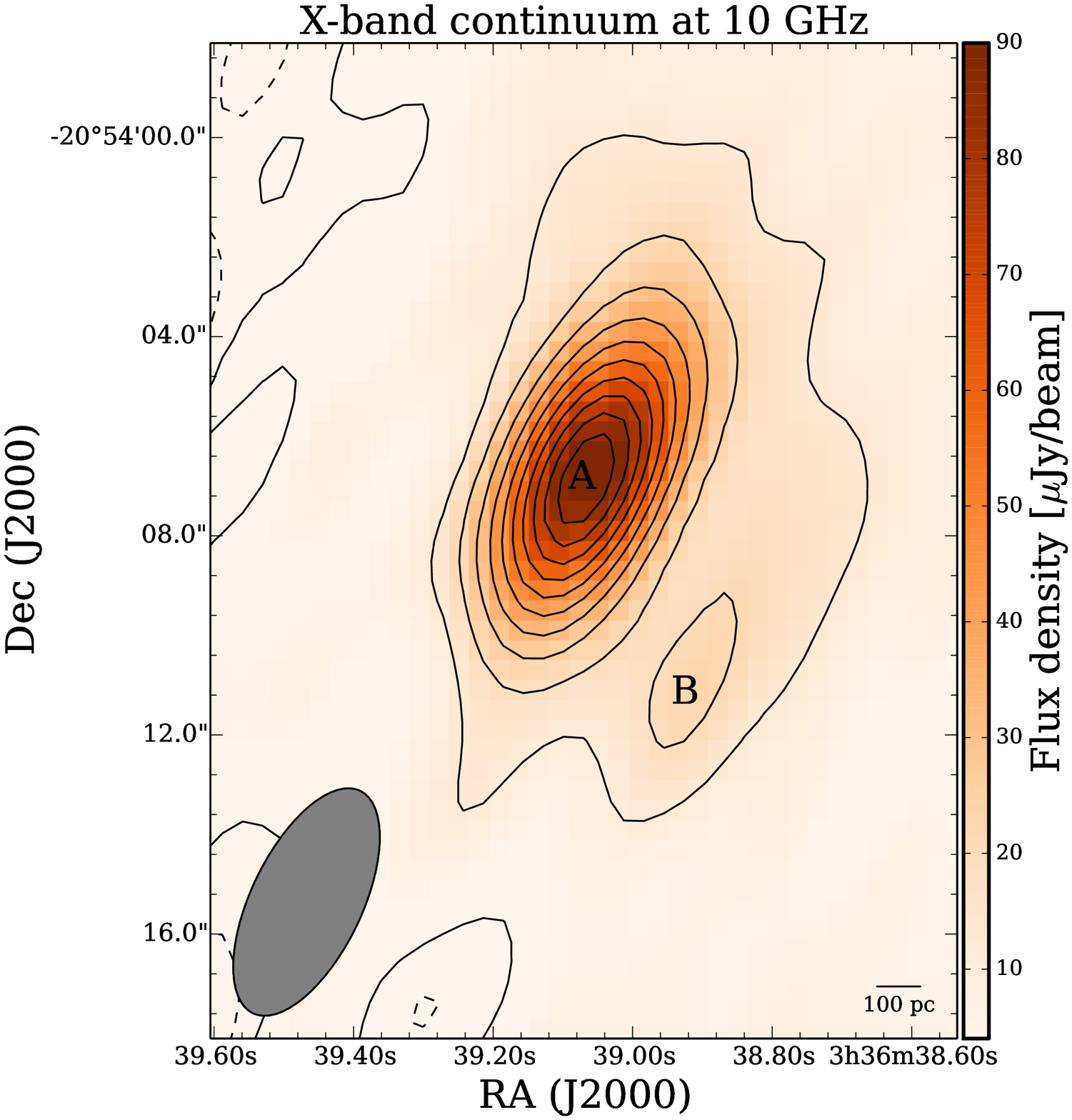}\hfill
\includegraphics[height=.5\textwidth,keepaspectratio]{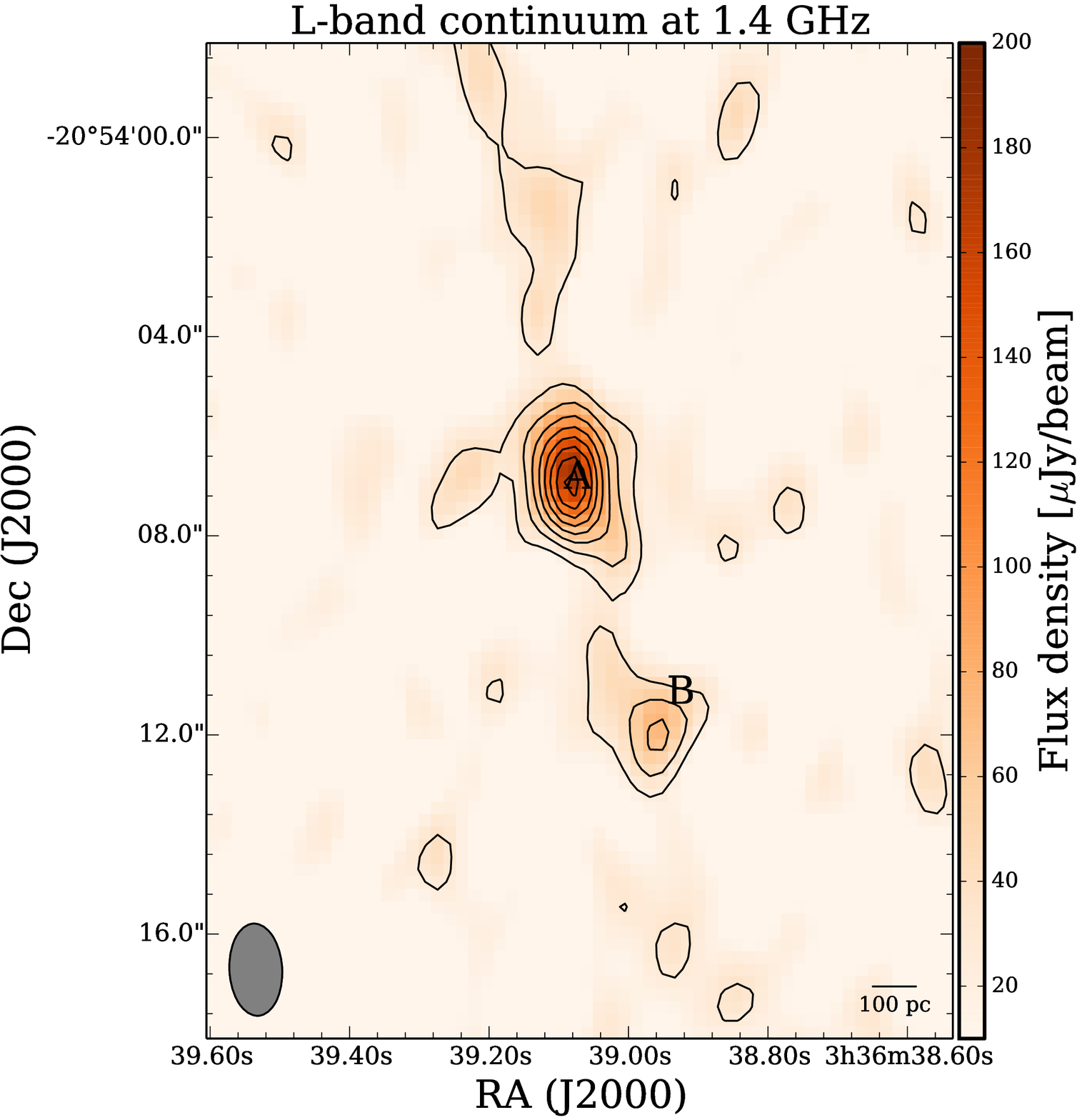}

\caption{ \label{fig:maps} Continuum emission maps at 10~GHz ({\it left}) and 1.5~GHz ({\it right}) obtained with the JVLA in NGC~1377. The solid contours are drawn every 2~$\sigma$ starting at 3~$\sigma$. The {\it rms} noise is 4~$\mu$Jy/beam and 13~$\mu$Jy/beam in the X- and L-band image, respectively. Negative values are drawn as dashed contours. The two main emission components are labeled as $A$ and $B$, see text for discussion.}
\end{center}
\end{figure*}

\begin{table*}
\begin{center}
\caption{\label{tab:journal} Journal of JVLA observations.}
\input{journal.tex}
\end{center}
\end{table*}

\subsection{Radio JVLA X-band}
We observed NGC\,1377 using the JVLA in C-configuration at X-band (8-12\,GHz) in full polarization mode, under project 14B-120. The total bandwidth was 4\,GHz, split in 32 spectral windows. The observations were performed in October 2014. The total on-source time was 1.8~hours. We used 3C\,138 for flux and bandwidth calibration, and J0340-2119 as the phase calibrator, at an angular distance of $1.0^{\circ}$ from NGC\,1377.

We used the Common Astronomy Software Applications package \citep[CASA, ][]{mcmullin2007} to perform a standard data reduction. We imaged the data using the multi-frequency synthesis (MFS) algorithm with a natural weighting and a pixel size of $0.4''$. The synthesized beam size was $5.0''\times2.2''$, with a position angle (PA) of $-26^{\circ}$. We achieved a $rms$ noise level of 4\,$\mu$Jy/beam. The resulting map is shown in Fig. \ref{fig:maps}.

\subsection{Radio JVLA L-band}
\label{sec:lobs}

\begin{figure*}
\includegraphics[width=.9\textwidth,keepaspectratio]{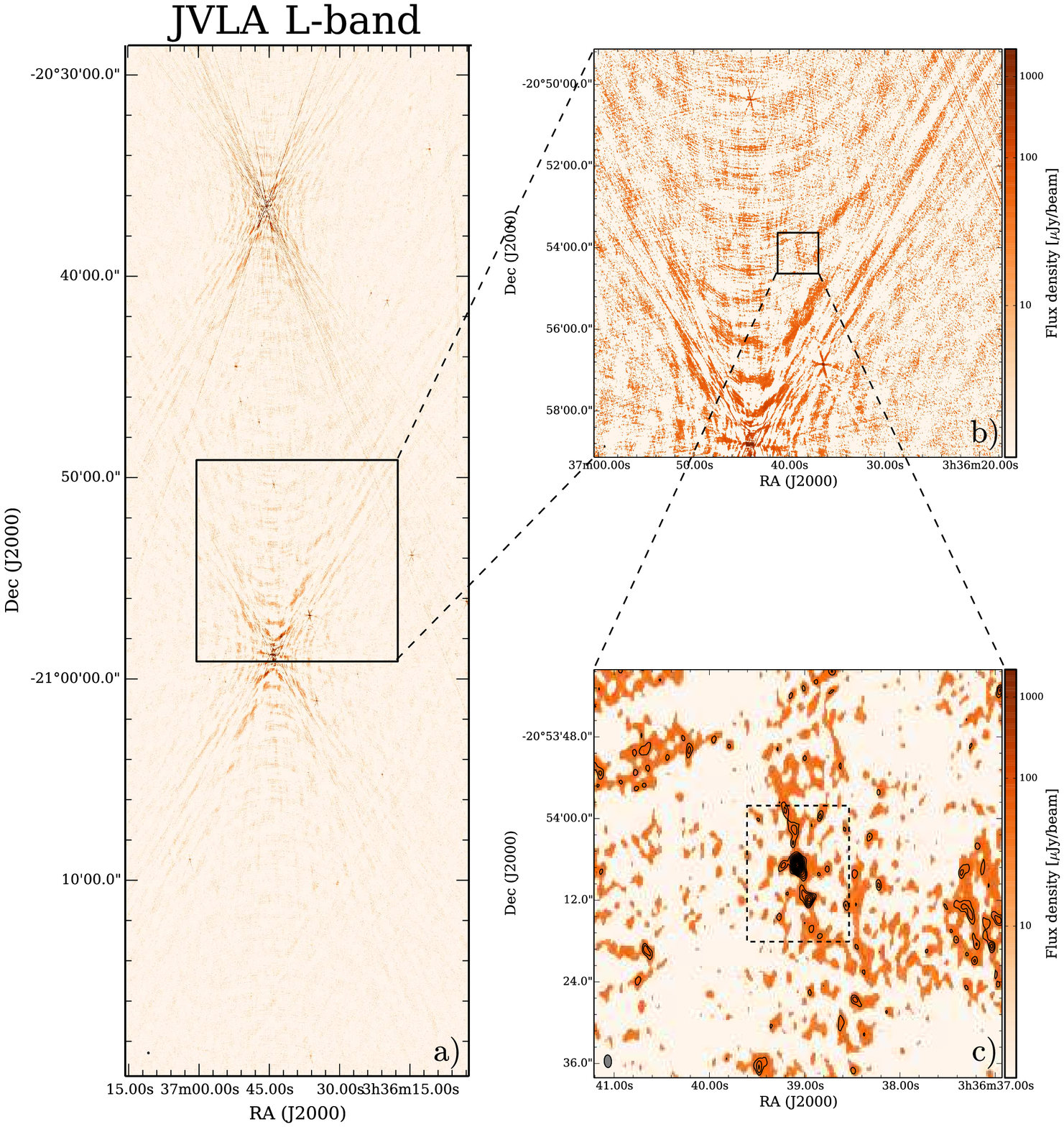}
\caption{\label{fig:lfov} {\it a) } Large-scale clean map of the JVLA L-band observations. The strongest source in the field is NVSS\,J033644-205849 (South of NGC~1377), with a peak flux of 64.7~mJy. The image has a dynamic range of 6470, with an {\it rms} of 10~$\mu$Jy. {\it b) } $10'\times10'$ region around NGC~1377 showing the main sources of contamination. {\it c) } $1'\times1'$ region around NGC~1377. Contours are drawn every $\sigma$, starting at 3--$\sigma$. The dashed rectangle marks the area shown in Figures~\ref{fig:maps}, \ref{fig:xmap}, \ref{fig:spind}, and \ref{fig:comap}.}
\end{figure*}

L-band observations at 1.5 GHz were carried out in June 2015 (project 15A-501) with the JVLA in A-configuration, with a bandwidth of 1\,GHz. The data reduction was analogous to our X-band data. We imaged the data using a cell size of $0.25"$ and a synthesized beam size of $1.8''\times1.1''$, with a PA of $3^{\circ}$. The total on-source time was 2~hours.

The MFS imaging of the L-band data was complicated by the presence in the field of two strong radio sources: NVSS\,J033644-205849 located 5\,arcmin to the South of NGC\,1377 and with a peak flux density of $65\,\mathrm{mJy\,beam}^{-1}$, and NVSS\,J033645-203637, 16\,arcmin to the North, with $14\,\mathrm{mJy\,beam}^{-1}$. This complication was twofold: first, the dynamic range of the image was limited, not being able to reach the nominal rms, and second, intense side-lobes appeared when we only imaged the area around NGC\,1377. This was not problematic in the X-band observations because of the steep spectral index of the two bright sources. To minimize the side-lobes, we mapped a very large area ($50\times17$\,arcmin) to include both sources in the field of view (see Fig.~\ref{fig:lfov}). While at such angular distances, smearing effects appear, the side-lobes of the resulting image were significantly reduced. We finally performed several phase-only self-calibration rounds, to improve the calibration of the data, yielding to a local off-source rms of $10\,\mu\mathrm{Jy\,beam}^{-1}$. The resulting L-band continuum image is shown in Fig. \ref{fig:maps}, while a journal of the observations is reported in Table~\ref{tab:journal}.

\subsection{Chandra X-rays}
\label{sec:xobs}
\textit{Chandra} observed NGC\,1377 on 2013-12-10 for 49 ks with ACIS-S (ObsID: 16086). The data reduction was performed using the CIAO 4.7 software. To obtain an image and spectra, we followed the \textit{Chandra} analysis guide \footnote{http://cxc.harvard.edu/ciao/guides/\#acis} and first run the \texttt{chandra\_repro} script to produce cleaned {\it level=2} event files. 

\begin{figure}
\centering
\includegraphics[height=.5\textwidth,keepaspectratio]{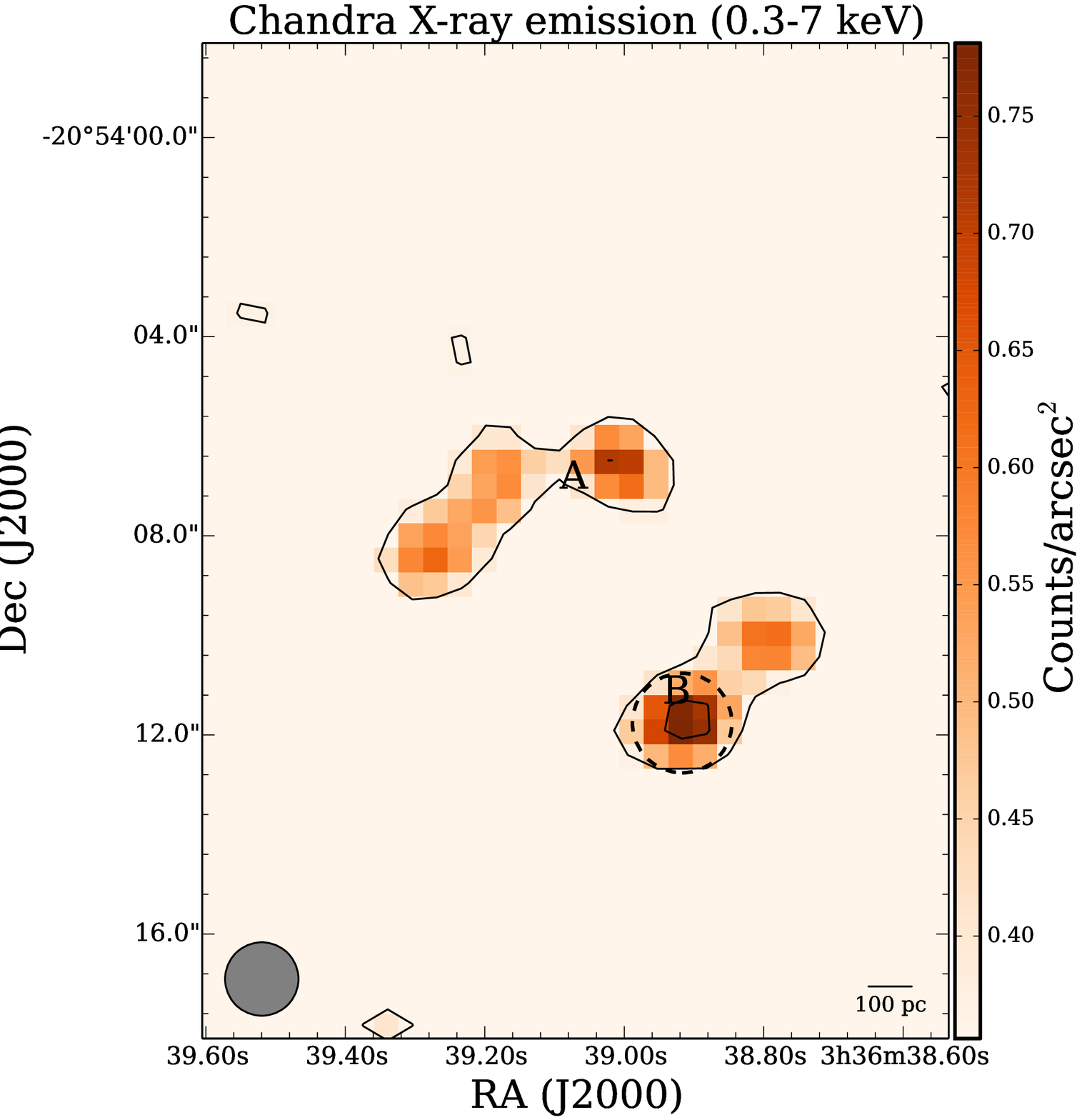}
\caption{\label{fig:xmap} Detection of X-rays emission with \textit{Chandra}. The observations were smoothed to a resolution of 1$''$.5. Contours are drawn every 0.35 counts/arcsec$^2$ starting at 0.35 counts/arcsec$^2$. The galaxy center is marked as A. The dashed ellipse shows the 2$''\times$2$''$ region around position B which was used to calculate a signal-to-noise of 2.2-$\sigma$ (see Section~\ref{sec:xobs} for discussion).}
\end{figure}

The resulting image is shown in Fig.~\ref{fig:xmap}. No emission is detected at the known position of the center of NGC\,1377, but a faint signal is present to the South-West, about 4$''$.5 from the center. To estimate the significance of this detection we integrated the X-ray emission in a circular aperture of 2$''$ centered at 3h36m38.93s,-20$^\circ$54$'$10.9$''$. As background region we chose an annulus centered on the source position with an inner radius of 7$''$ and an outer radius of 14$''$. In the source region we detect 5 counts, which by applying Poisson statistics correspond to a signal-to-noise ratio of 2.2. This detection has thus to be considered as tentative. A similar analysis was performed for the faint emission around the galaxy center which has signal-to-noise ratio lower than 2 and will be considered as a non-detection. 



\section{Results}
\label{sec:res}
\subsection{Morphology of the radio continuum}

The radio continuum of NGC~1377 is detected, for the first time, in both X and L bands with peak flux densities of 82$\pm$4~$\mu$Jy/beam and 178$\pm$13~$\mu$Jy/beam, respectively. The maps in Fig.~\ref{fig:maps} show that in both bands the emission is distributed in two main components, labeled A and B. The A component contains most of the flux and is peaked on the center of the galaxy, while the B component is displaced to the South-West by 4$''$.5, corresponding to $\sim$500~pc at the galaxy's distance. 

The B component is detected with a signal to noise ratio of 7 in X-band and 5 in L-band, which makes it unlikely for the emission to be completely caused by residual effects from the synthesized beam.  We note however that the position of the peak of the B component is not exactly the same in the L- and X-band images, which may be due to a residual effect from the strong steep-spectrum source in the field of view of L-band observations (see Section~\ref{sec:lobs}). With the data at our disposal we cannot exclude that a fraction of the emission of the L-band B component is due to un-removed emission from such a strong radio source, which escaped the MFS cleaning.

To better characterize the structure of the radio emission we performed a 2D Gaussian fit with the CASA routine {\tt imfit}. The results of the fit are shown in Table~\ref{tab:gaussfit}. In X-band, the A component has a flux density of 160$\pm$12~$\mu$Jy and is barely resolved, with a de-convolved size of 4$''$.4$\times$2$''$.5, which is only slightly larger than the synthesized beam. The B component is best fit by a point source of flux density 25$\pm$2~$\mu$Jy. A similar result is found in L-band, where the A component is barely resolved, with a de-convolved size of 2$''$.6$\times$1$''$.7 and flux density 427$\pm$69~$\mu$Jy. The B component in L-band is best fit by a point source of 101$\pm$19~$\mu$Jy.

\begin{table*}
\begin{center}
\caption{\label{tab:gaussfit} Results of the Gaussian fit of the continuum components in X- and L-band with the CASA routine {\tt imfit}.}
\input{source_fit.tex}
\end{center}
\end{table*}

\subsection{Radio spectral index}
\begin{figure*}
\begin{center}
\includegraphics[height=.5\textwidth,keepaspectratio]{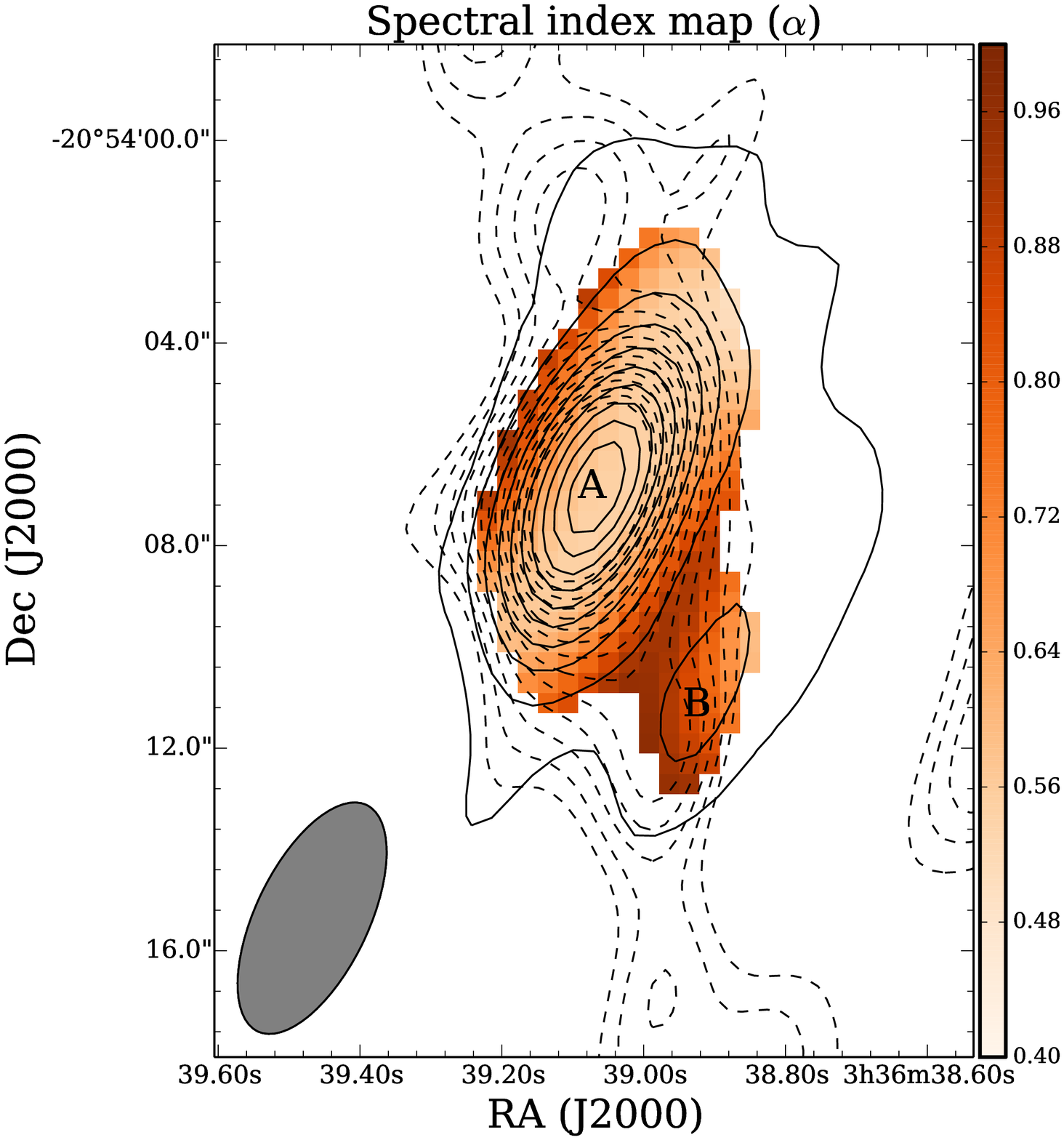}
\includegraphics[height=.5\textwidth,keepaspectratio]{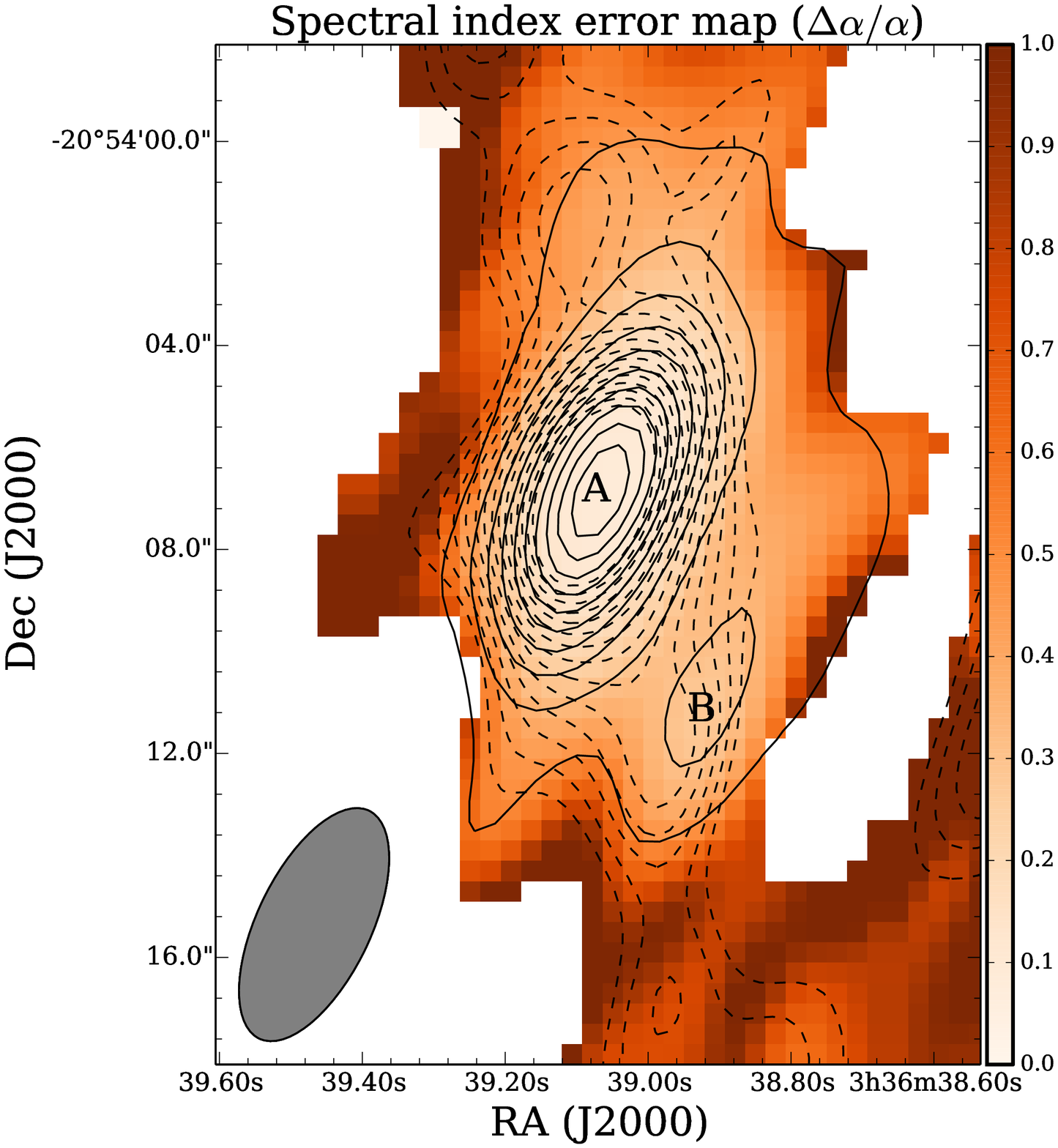}
\caption{\label{fig:spind} {\it Left:} Spectral index map for the radio emission in NGC~1377, derived from X- and L-band observations. Values are only shown where the spectral index uncertainty $\Delta\alpha/\alpha$ is less than 30$\%$. Solid contours show the X-band emission and are the same as in Fig.~\ref{fig:maps}. Dashed contours show the L-band continuum emission smoothed to the X-band resolution, see text for discussion. The two main emission components are labeled as $A$ and $B$, as in Fig.~\ref{fig:maps}. {\it Right:} Map of relative uncertainty for the derived spectral index.}
\end{center}
\end{figure*}

Assuming the radio spectral energy distribution (SED) to be described by power law of the kind $S_\nu\propto\nu^{-\alpha}$, the Gaussian fit of Table~\ref{tab:gaussfit} results in a spectral index $\alpha$=0.5$\pm$0.1 for the A component and $\alpha$=0.7$\pm$0.2 for the B component.

In order to produce a map of the radio spectral index in NGC~1377, a low-resolution L-band map was derived by smoothing the L-band clean image to the X-band resolution with the CASA task {\tt imsmooth}. This allowed  a pixel-to-pixel comparison of the intensities in the X- and L-band maps. The resulting spectral index map is shown in Fig.~\ref{fig:spind}. The map shows a spectral index gradient, with $\alpha$ steepening to the South-West from a value of 0.6 at the A position to 0.7--0.8 for the B component. 

The spectral index derived for the B component is very close to the non-thermal spectral index observed in star forming galaxies and AGN radio jets, which is generally associated with optically thin synchrotron emission ($\alpha\simeq0.7-0.8$, e.g \citet{condon92}). However, because of a possible contribution of the dirty beam from the steep spectrum source in the L-band (see Section~\ref{sec:lobs}), the spectral index of 0.7 for the B component has to be considered as an upper limit.

\subsection{Constraints on the free-free opacity}

\begin{figure*}
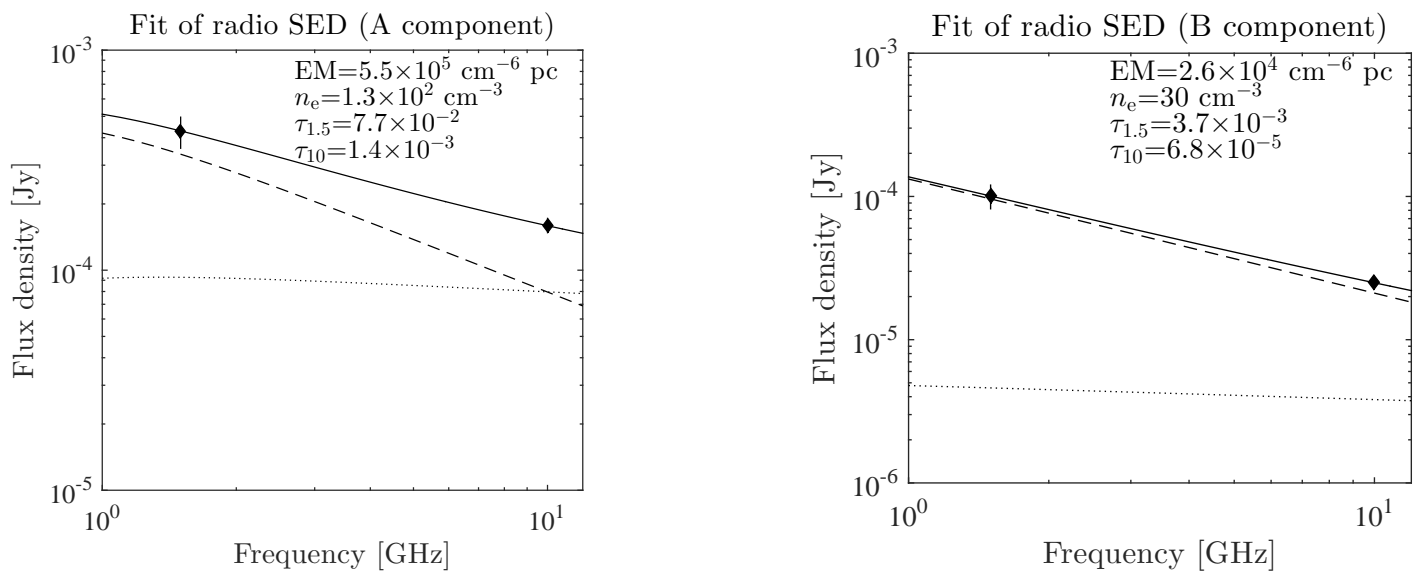

\begin{center}
\includegraphics[height=.4\textwidth,keepaspectratio]{ff_sedA.eps}\hfill
\includegraphics[height=.4\textwidth,keepaspectratio]{ff_sedB.eps}
\caption{\label{fig:sedfit} Fit of the synchrotron and thermal free-free component of the radio emission of NGC~1377 for the A ({\it left}) and B ({\it right}) components. The measured flux density at L- and X-band is shown as filled diamonds. The total fit is shown as a {\it solid} line. The synchrotron and free-free components are drawn as a {\it dashed} and {\it dotted} line, respectively.}
\end{center}
\end{figure*}

\begin{figure*}
\begin{center}
\includegraphics[height=.4\textwidth,keepaspectratio]{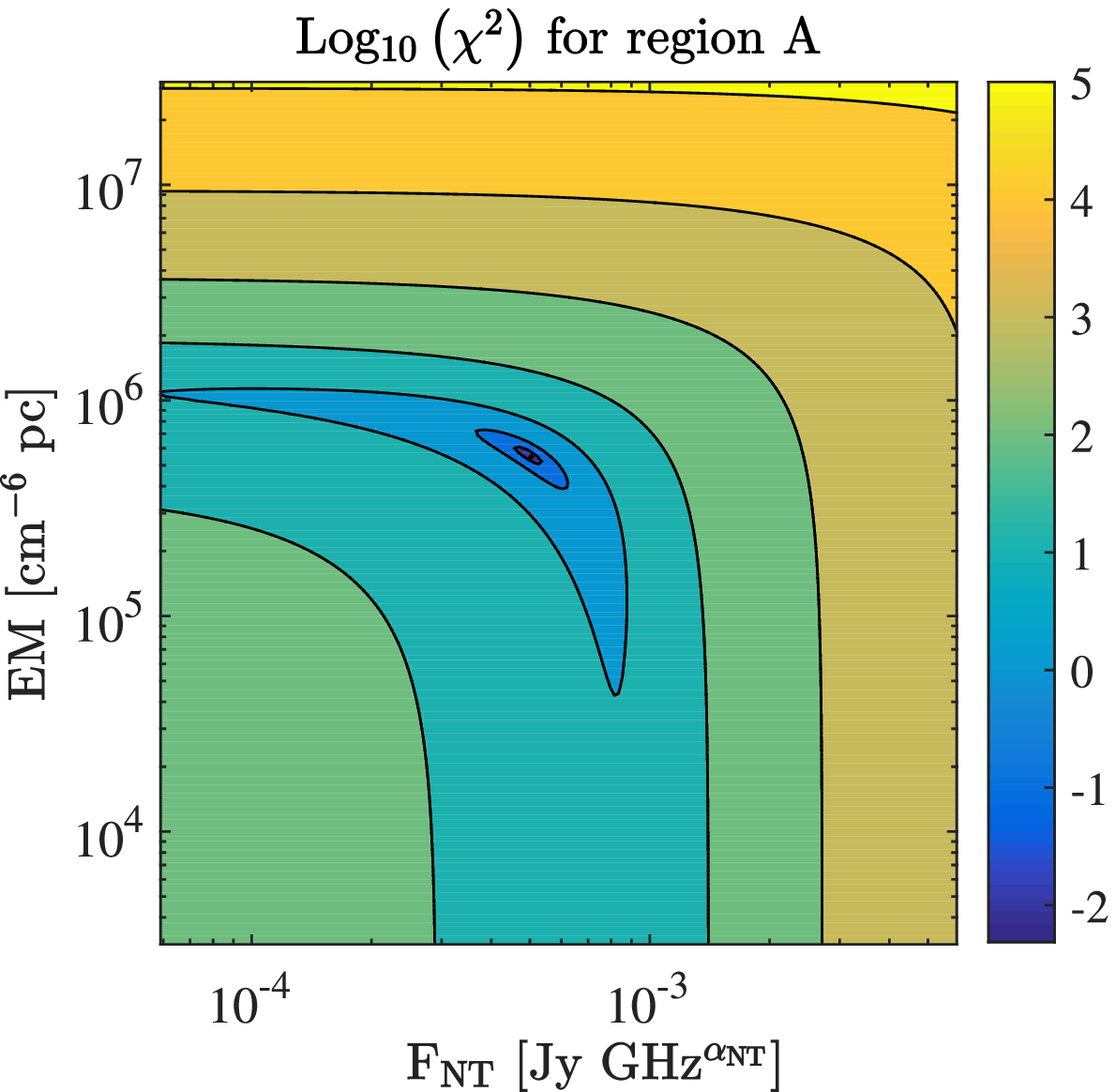}\hfill
\includegraphics[height=.4\textwidth,keepaspectratio]{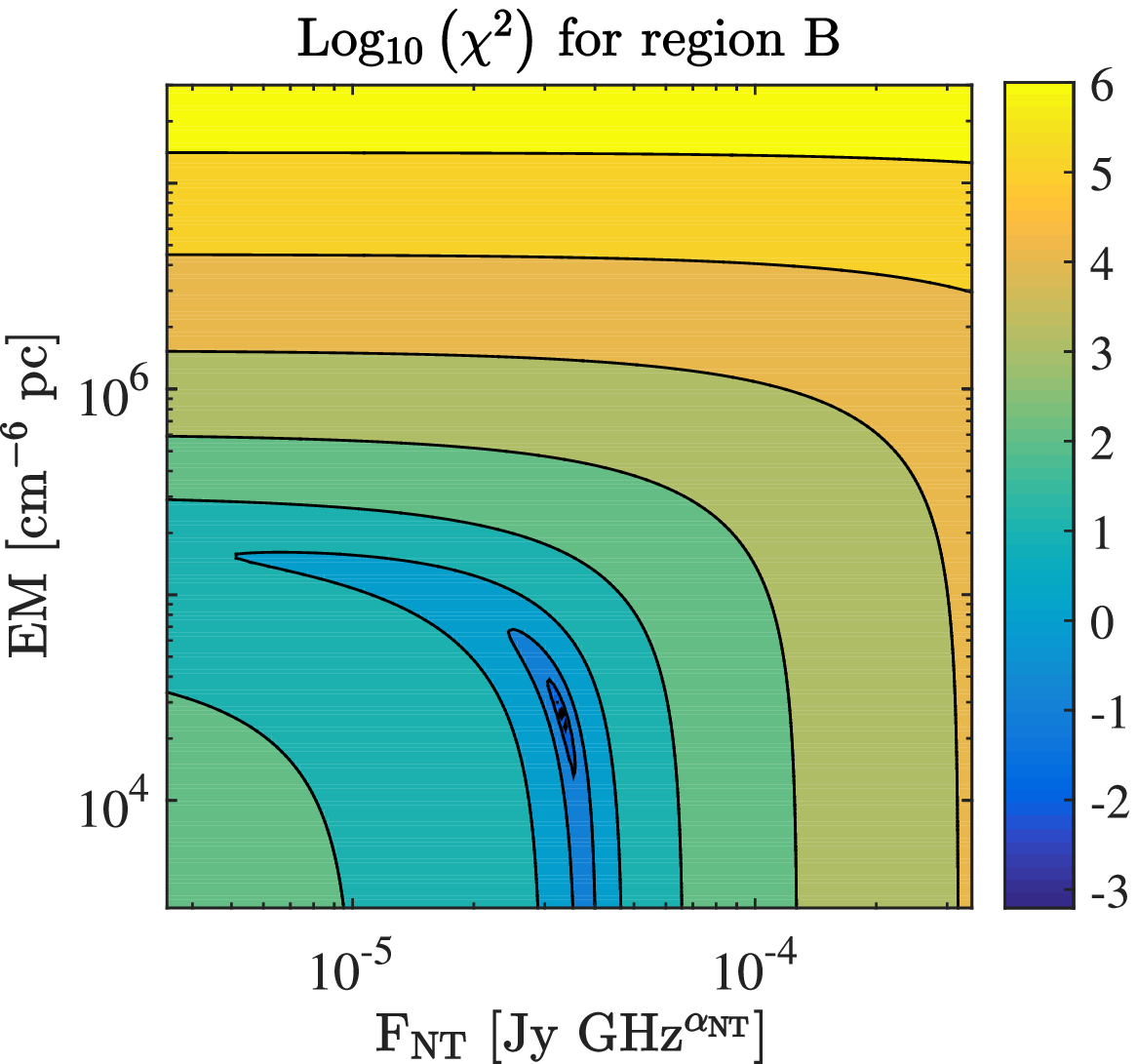}
\caption{\label{fig:chisurf} $\chi^2$ surfaces for the fit of the radio spectral energy distribution of the A ({\it left}) and B ({\it right}) components. The synchrotron constant is defined as F$_\mathrm{NT}\equiv$S$_\mathrm{NT}$(1.5 GHz)/1.5$^{-\alpha_\mathrm{NT}}$, where S$_\mathrm{NT}$(1.5 GHz) is the synchrotron flux at 1.5~GHz and $\alpha_\mathrm{NT}$ is the non-thermal spectral index.}
\end{center}
\end{figure*}

\label{sec:freefree}
One of the possible explanations for the extreme radio deficiency of NGC~1377 has been proposed to be free-free absorption of the synchrotron emission from a compact starburst or AGN \citep[e.g.,][]{roussel03,imanishi09}. Our detection of the X- and L-band fluxes allow us to put some constraints on the free-free opacity and electron density in the galaxy.

If we assume the radio spectrum to be due to a synchrotron source and foreground thermal free free emission, we can write the total radio flux density as 
\begin{equation}\label{eq:stot}
S=S_\mathrm{NT}\,e^{-\tau_\mathrm{T}}+S_\mathrm{T}\left(1-e^{-\tau_\mathrm{T}}\right),
\end{equation}
where $S_\mathrm{NT}$ is the non-thermal synchrotron emission, $S_\mathrm{T}$ is the thermal free-free emission, and $\tau_\mathrm{T}$ is the free-free opacity. Following \citet{condon92}, 
\begin{equation}
\tau_\mathrm{T}=3.3\times10^{-7}\left(\frac{EM}{\mathrm{cm^{-6}~pc}}\right)\left(\frac{T_\mathrm{e}}{10^4~\mathrm{K}}\right)^{-1.35}\left(\frac{\nu}{\mathrm{GHz}}\right)^{-2.1},
\end{equation}

where $T_\mathrm{e}$ is the electron temperature and $EM=\int n_\mathrm{e}^2~dl$ is the emission measurement determined by the electron density $n_\mathrm{e}$ and the extent of the free-free absorber $dl$ along the line of sight. In our case we assume a constant $n_\mathrm{e}$ in a spherical region with diameter 2$''$ ($\sim$230~pc), which is the angular size determined by the Gaussian fit of Table~\ref{tab:gaussfit} for the A component. For the B component, the chosen value is an upper limit to the size of the emitting region.

We assume the thermal free-free emission to be a black body at temperature $T_\mathrm{e}$ in the Rayleigh–Jeans regime:
\begin{equation}
S_\mathrm{T}=\Omega\frac{k_\mathrm{B}T_\mathrm{e}\nu^2}{\mathrm{c}^2},
\end{equation}
with $\Omega$ the solid angle subtended by the source. For the non-thermal emission we assume a form $S=S_\mathrm{NT}=F_\mathrm{NT}\nu^{-\alpha_\mathrm{NT}}$, with $F_\mathrm{NT}$ constant. 

We assume an optically-thin synchrotron spectral index of $\alpha_\mathrm{NT}=$0.8 and an electron temperature $T_\mathrm{e}=10^4$~K, as in  \citet[][]{condon92}. At low opacities, the thermal part of Eq.~\ref{eq:stot} has only a very weak dependence on the electron temperature ($S_\mathrm{T}\tau_\mathrm{T}\propto T_\mathrm{e}^{-0.35}$), so that an error of one order of magnitude in $T_\mathrm{e}$ would change the resulting free-free emission by only a factor of two. With our assumptions, the observed radio flux density only depends on the value of the non-thermal constant $F_\mathrm{NT}$ and on the electron density $n_\mathrm{e}$, which can be determined by fitting Eq.~\ref{eq:stot} to the L- and X-band fluxes. The results of the $\chi^2$-minimization fit are shown in Fig.~\ref{fig:sedfit}, while the $\chi^2$-surfaces are shown in Fig.~\ref{fig:chisurf}. 

For the A component we find that the emission is best fit by $EM\simeq$10$^4$~cm$^{-6}$~pc, corresponding to an electron density $n_\mathrm{e}\simeq$7~cm$^{-3}$. The free-free opacity is very low, of the order of 10$^{-3}$ and 10$^{-5}$ at 1.5 and 10~GHz, respectively. The free-free contribution to the total radio flux goes from 25\% at 1.5 GHz to about 50\% at 10 GHz. For the B component the fit results in an $EM<$10$^3$~cm$^{-6}$~pc, corresponding to an electron density $n_\mathrm{e}<$2~cm$^{-3}$. The free-free opacity in both L- and X-band is found to be negligible, with values $<$10$^{-5}$. 
The free-free contribution to the total radio flux goes from 5\% at 1.5 GHz to about 15\% at 10 GHz. 

These results should be considered as order-of-magnitude estimates because of the many assumptions of our analysis. The free-free contribution resulting from our fit depends on the assumed non-thermal spectral index, which in some cases (e.g., in AGN) could be lower than the standard $\alpha_\mathrm{NT}=$0.8. In our analysis, a shallower non-thermal spectral index would result in an even lower free-free emission. The estimated free-free contributions have thus to be considered as upper limits. The main uncertainty affecting the derived electron densities comes from the unknown size of the free-free emission region. For the A component the choice of a 2$''$ source is justified by our Gaussian fitting of Table~\ref{tab:gaussfit} and partly by molecular gas observations. \citet{aalto1377} found that the CO~2--1 emission is concentrated in the inner 200~pc of the galaxy and we can assume most of the free-free to be associated with star formation taking place in this molecular zone. However, it is unclear how much of the observed CO emission is associated with the molecular disk and how much is instead part of a molecular outflow. Recent results by \citet{aalto2016} suggest that the launching region of the molecular outflow could be smaller than 30~pc and that the energy source should be located inside this region. A diameter of 30~pc for the radio source would imply an electron density $n_\mathrm{e}\simeq$10$^2$~cm$^{-3}$, corresponding to an L-band opacity $\tau_\mathrm{T}$=0.08.

\subsection{The radio-infrared relation}\label{sec:rfir} One of the most used calibrations in extragalactic star-formation is the observed correlation of radio and infrared fluxes \citep[e.g., ][]{helou_85,condon92,yun2001}, described by the factor
\begin{equation}
\label{eq:q}
q_\mathrm{IR}\equiv log_{10}\left(\frac{L_\mathrm{IR}}{3.75 \times 10^{12} L(1.4~\mathrm{GHz}) } \right).
\end{equation}
The observed $q_\mathrm{IR}$ varies from 2.34$\pm$0.26~dex \citep{yun2001} to  2.64$\pm$0.26~dex \citep{bell2003}, depending on whether L$_\mathrm{IR}$ is integrated over the far-IR or whole IR range. The tight correlation observed for large samples of star-forming galaxies can be explained by both the IR and radio emission to be tracing massive star formation, the IR tracing heating of dust by the UV from young stars and the radio tracing synchrotron emission from supernova-remnants \citep[e.g., ][]{condon92}. Only a few \% of the galaxies in the IRAS~2~Jy sample \citep{yun2001} deviate significantly from the radio-IR correlation. Radio-excess galaxies are generally associated with radio-loud AGN, while IR-excess galaxies have been suggested to harbor either a compact young starburst or an heavily obscured AGN and their nature is still debated \citep[e.g., ][]{roussel03,roussel06,costagliola2013}. 

For NGC~1377, \citet{imanishi09} find an IR luminosity of 4.8$\times$10$^{43}$~erg~s$^{-1}$~(1.2$\times$10$^{10}$~L$_\odot$) by integrating the IRAS fluxes from 12~$\mu$m to 100~$\mu$m. For the A component we find an L-band luminosity of 2.8$\times$10$^{26}$~erg~s$^{-1}$~Hz$^{-1}$, which for Eq.~\ref{eq:q} results into a $q_\mathrm{IR}\simeq$4.7. This value is about 8-$\sigma$ above the mean value of 2.64$\pm$0.26 found for most star-forming galaxies \citep[e.g., ][]{bell2003}. A similar result is found when using the original definition of $q_\mathrm{FIR}$ by \citet{helou_85} using the far-IR flux. For a L$_\mathrm{FIR}=$4$\times$10$^{9}$~L$_\odot$~\citep{roussel03} we find $q_\mathrm{FIR}$=4.2, which deviates for more than 7-$\sigma$ from the mean value of 2.34$\pm$0.26 found by \citet{roussel03}. Our measurement confirms the extreme IR-excess of NGC~1377, which was first discussed by \citet{roussel03}.

\subsection{Star formation rate indicators}
\label{sec:sfrind}
In this section we compare the SFR estimates derived from  different diagnostics and briefly discuss the implications for the properties of star formation in NGC~1377. Most of these methods have been discussed by \citet{murphy2011}, and by \citet{kennicutt2012}. The main results of our analysis are shown in Table~\ref{tab:sfr}.

{\it Optical tracers:} Optical spectroscopic observations by \citet{roussel06} find no H$\alpha$ emission in NGC~1377, but only weak NII and SII lines. These authors suggest that H$\alpha$ emission form a nuclear starburst may be strongly attenuated by dust, as supported by the large opacity of the central IR source, and that NII and SII emission would be emerging from shocked low-ionization foreground gas. The large obscuration towards the nucleus thus makes it impossible to use standard optical SFR estimators.

{\it Infrared tracers:} Radiation at infrared wavelengths can be used as a star formation tracer in highly obscured environments, assuming that most of the IR flux is generated by re-processed starlight. The emission at 24$\mu$m is produced by absorption by dust of the UV light from young massive stars and is thus a good tracer of recent star formation. By using the calibration by \citet{murphy2011} the 24~$\mu$m flux of NGC~1377 would correspond to a star formation rate of SFR$_{24\mu\mathrm{m}}$=2.6~M$_\odot$~yr$^{-1}$. A similar result is found for the SFR inferred from the total IR luminosity, SFR$_\mathrm{IR}$=1.9~M$_\odot$~yr$^{-1}$~(see Table~\ref{tab:sfr}, Eq. {\it h, i}).

{\it Radio tracers:} Radio emission from star-forming galaxies is generally composed of optically thin synchrotron emission from electrons accelerated by SN explosions in the galactic magnetic field and by thermal free-free emission from the ionized gas around massive stars. The synchrotron flux is directly linked to the supernova rate and can be translated into a SFR by assuming an initial mass function \citep[IMF, e.g., ][]{condon92}. For a Kroupa IMF \citep{kroupa01} the relation between non-thermal synchrotron flux and SFR found by \citet{murphy2011} reads
\begin{equation}
\mathrm{SFR_{NT}=6.64\times 10^{-29} \nu^{\alpha_{NT}} L_{NT}}, 
\end{equation}
which for NGC~1377 results in a SFR$_\mathrm{NT}$=0.02~M$_\odot$~yr$^{-1}$.

The thermal free-free emission depends on the hydrogen ionization rate by UV photons from massive young stars, which for continuous star formation is proportional to the SFR. \citet{murphy2011} find the following calibration 
\begin{equation}
\label{eq:sfrt}
\mathrm{SFR_T=4.6\times 10^{-28}\left(T_e/10^4~K\right)^{-0.45}\nu^{0.1} L_T},
\end{equation}
where T$_\mathrm{e}\simeq$10$^4$~K is the electron temperature and L$_\mathrm{T}$ is the thermal free-free luminosity in erg~s$^{-1}$~Hz$^{-1}$. Our best estimate of the thermal luminosity comes from the fit in Section~\ref{sec:freefree} of the radio continuum for the A component, giving L$_\mathrm{T}$(1.5~GHz)=7.1$\times$10$^{25}$~erg~s$^{-1}$~Hz$^{-1}$. Equation~(\ref{eq:sfrt}) then results in a SFR$_\mathrm{T}$=0.03~M$_\odot$~yr$^{-1}$, which is similar to what found from the non-thermal component.

{\it Schmidt-Kennicutt relation:} A correlation of the form $\Sigma_\mathrm{SFR}\propto\Sigma_\mathrm{gas}^{\alpha_\mathrm{SK}}$ between the surface gas density and star formation rate density in the disks of galaxies is observed throughout the Universe and is often referred to as the Schmidt-Kennicutt (SK) relation \citep[e.g., ][]{kennicutt2012}. The exponent $\alpha_\mathrm{SK}$ varies in the range 1-1.5, depending on the gas tracer used and on the galaxy type \citep[e.g., ][]{genzel2010,daddi2010}, but for local spirals the star formation and molecular gas surface densities are well described by the relation found by \citet{kennicutt98}
\begin{equation}
\label{eq:ken98}
\Sigma_\mathrm{SFR}=(2.5\pm0.7)\times 10^{-4}\Sigma_\mathrm{gas}^{1.4\pm0.15}~\mathrm{M_\odot~yr^{-1}~kpc^{-2}},
\end{equation}
where $\Sigma_\mathrm{gas}$ is the molecular gas surface density in M$_\odot$~pc$^{-2}$. 

Interferometric observations of CO emission with the SMA \citep{aalto1377} and ALMA \citep{aalto2016} find that the molecular gas is concentrated in the inner 200 pc of the galaxy in a disk-outflow system (see Fig.~\ref{fig:comap}). The dynamical mass of the inner 60~pc molecular disk is 4$\times$10$^7$~M$_\odot$, corresponding to a gas surface density $\Sigma_\mathrm{gas}$=1.4$\times$10$^{4}$~M$_\odot$~pc$^{-2}$. The correlation of Eq.~(\ref{eq:ken98}) would predict a star formation rate density  $\Sigma_\mathrm{SFR}\simeq$160~M$_\odot$~yr$^{-1}$~kpc$^{-2}$, corresponding to a SFR$_\mathrm{SK}$=0.45$\pm$0.12~M$_\odot$~yr$^{-1}$ integrated over the molecular disk (Eq.~{\it k} in Table~\ref{tab:sfr}).

\begin{table*}
\caption{\label{tab:sfr} Multi-wavelength estimates of the star formation rate in NGC~1377.}
\input{sfr_table.tex}
\end{table*}

\subsection{Starburst99 models}
\label{sec:sb99}

\begin{figure*}
\begin{center}
\includegraphics[height=.45\textwidth,keepaspectratio]{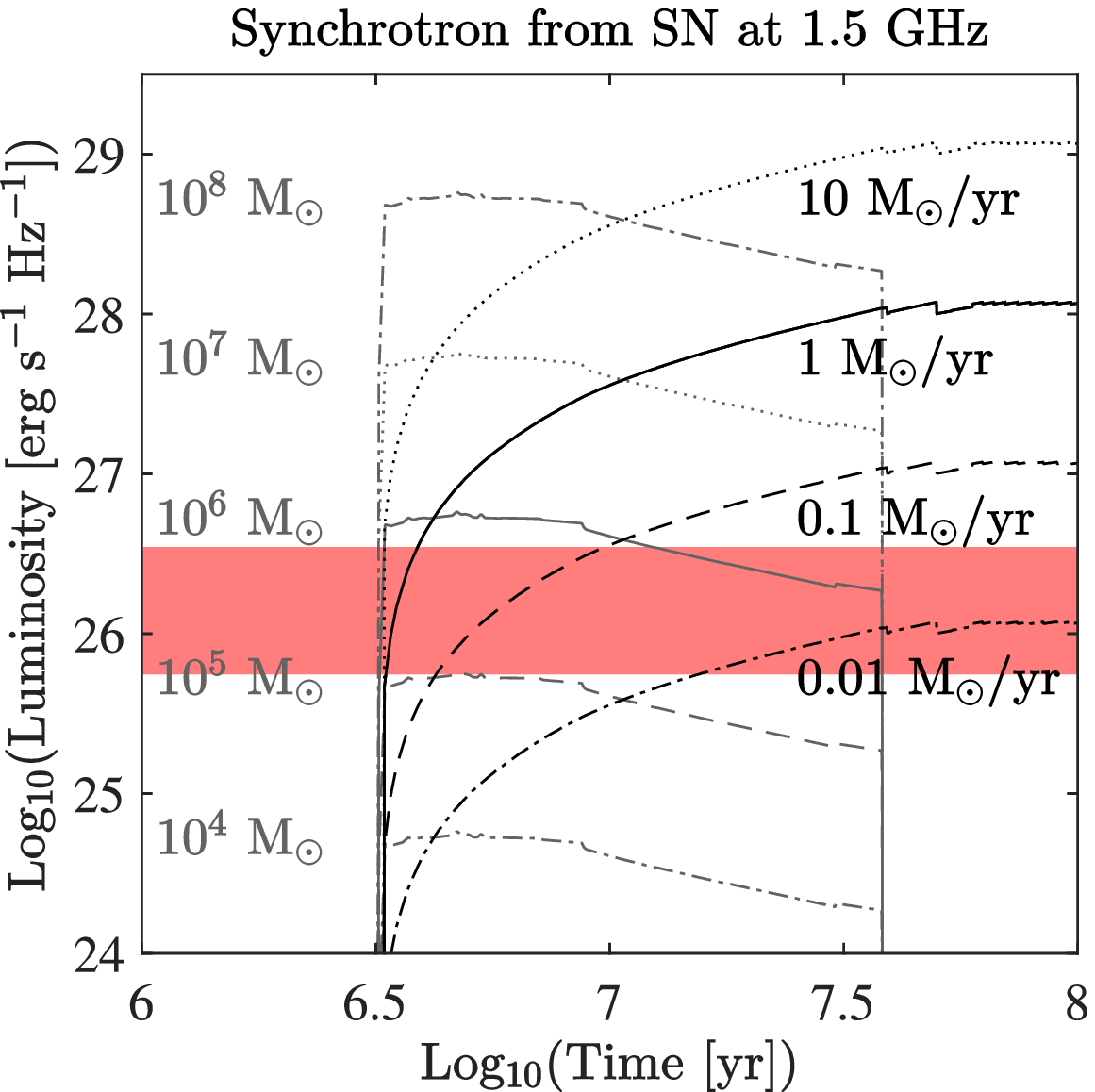}(a)\hfill
\includegraphics[height=.45\textwidth,keepaspectratio]{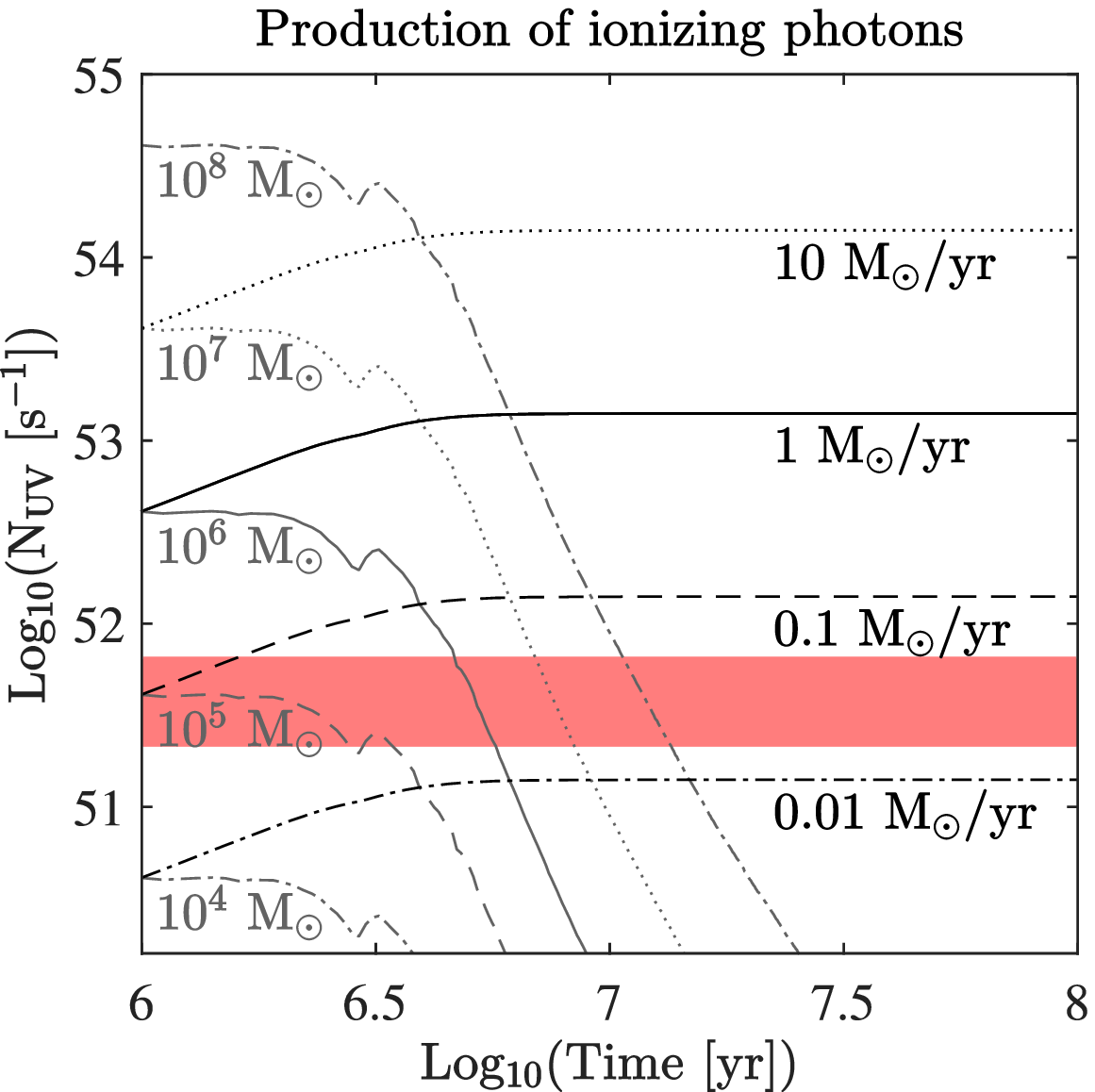}(b)\\
\vspace{10pt}
\includegraphics[height=.45\textwidth,keepaspectratio]{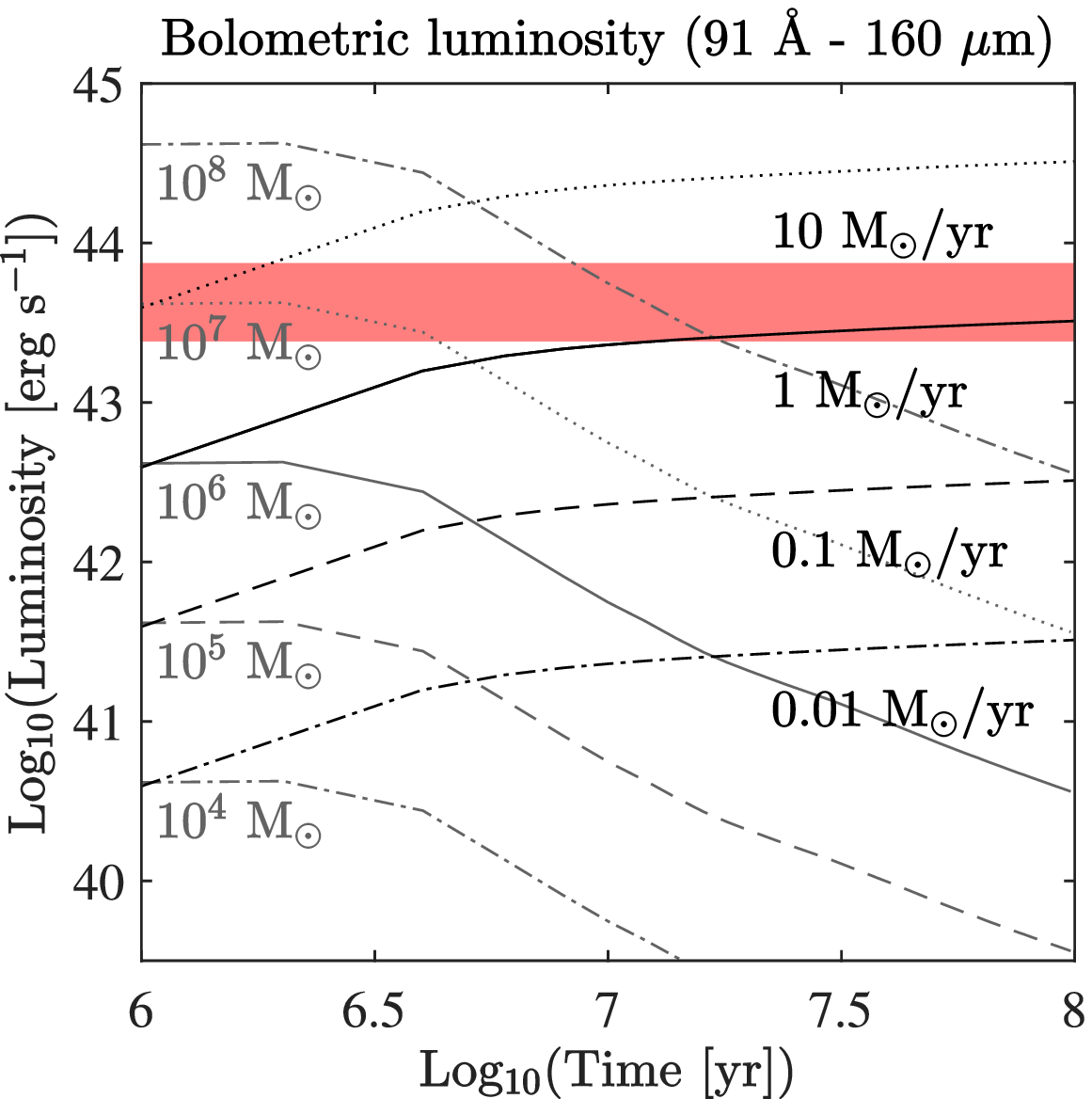}(c)\hfill
\includegraphics[height=.45\textwidth,keepaspectratio]{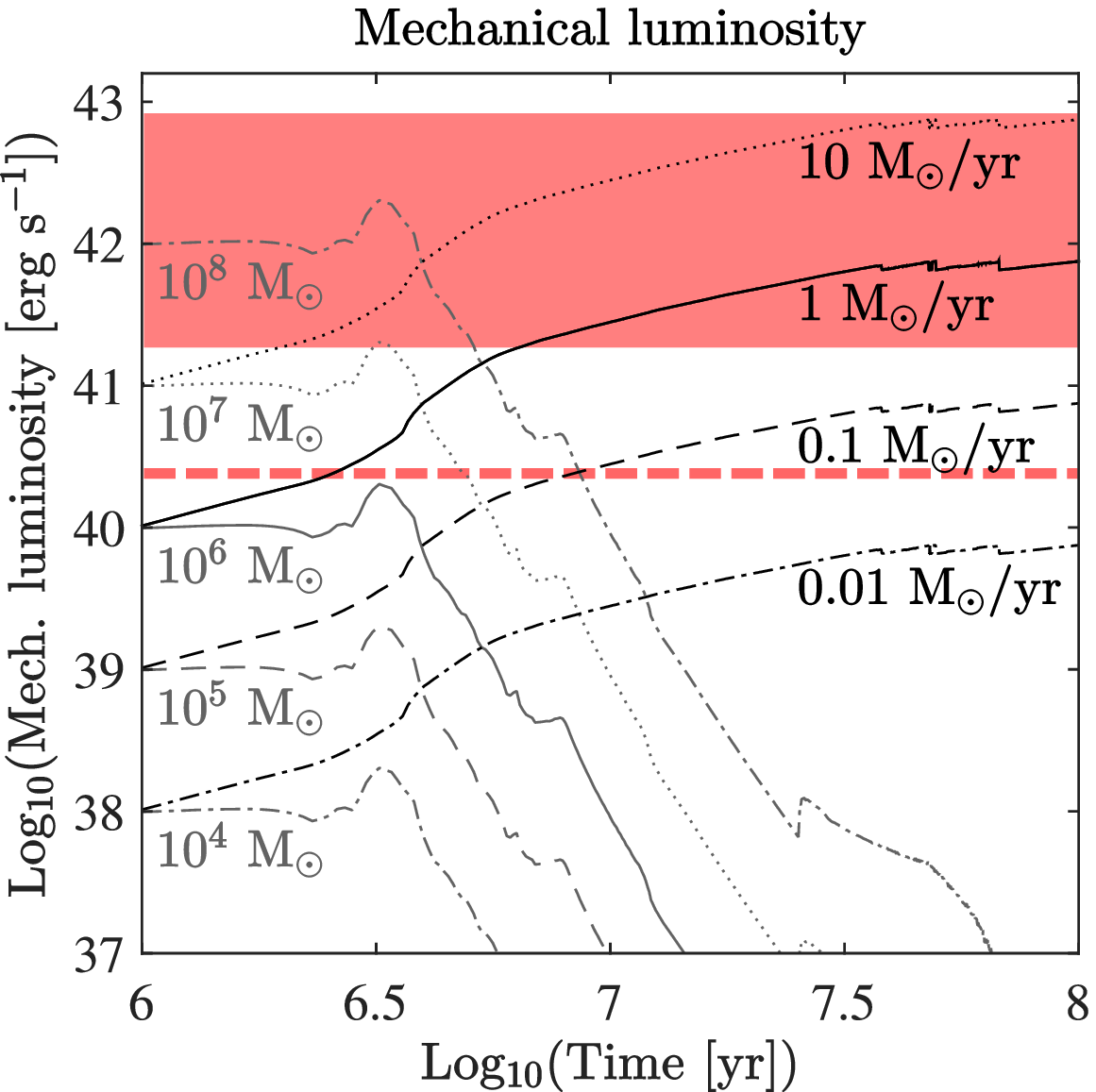}(d)

\caption{\label{fig:sb99} Comparison of observed properties of NGC~1377 with {\tt Starburst99} models. 
{\it Black lines} show results for models with continuous star formation rates of 0.01, 0.1, 1, and 10~M$_\odot$/yr. {\it Gray lines} show results for models assuming instantaneous star formation with total stellar masses ranging from 10$^4$ to 10$^8$~M$_\odot$. {\it (a):} Comparison of the observed X-band emission with the expected synchrotron emission at 1.5~GHz from supernovae. The shaded area shows the confidence interval of our measurement of the radio luminosity at 1.5~GHz. {\it (b):} Production rate of ionizing photons ($N_\mathrm{UV}$) from HII regions. The shaded area shows the confidence interval for $N_\mathrm{UV}$ derived from our estimate of the thermal radio emission. {\it (c): } Bolometric luminosity derived by integrating the continuum flux from the {\tt Starburst99} models between 91~\mbox{\AA}~and 160~$\mu$m. The shaded area shows the IR luminosity derived by the IRAS fluxes \citep{imanishi09} assuming a confidence interval of 50\%. {\it (d): }  Mechanical luminosity from stellar winds and supernova explosions. The  shaded area shows the confidence interval of the measurement of mechanical luminosity of the molecular outflow detected by \citet{aalto2016} with ALMA in the CO~J=3--2 transition, assuming a Galactic CO-H$_2$ conversion factor ({\it massive jet}). The red dashed line shows the lower limit of the outflow mechanical luminosity assuming a conversion factor 10 times lower ({\it light jet}).  See Section~\ref{sec:sb99} for discussion.}
\end{center}
\end{figure*}

The star-formation rate estimators discussed in Section~\ref{sec:sfrind} are summarized in Table~\ref{tab:sfr}. Because of the large uncertainties affecting these measurements \citep[e.g., ][]{kennicutt2012} the SFR found have to be considered as order-of-magnitude estimates. Nevertheless, a significant scatter is evident, with IR diagnostics resulting in SFR more than two orders of magnitude greater than those estimated from our radio continuum observations. The most used SFR calibrators assume a continuous star formation and starburst ages greater than 100~Myr \citep[e.g., ][]{murphy2011}, the observed discrepancy may be thus due to a time evolution of the observable properties of the star-forming region.

In previous studies, when only upper limits to the radio continuum of NGC~1377 were known, the extreme radio deficiency of NGC~1377 has been interpreted as the evidence of a nascent, pre-supernova starburst \citep[e.g., ][]{roussel06}. Synchrotron emission from supernovae is only produced a few Myr after the onset of the star-forming event, because massive stars need time to explode into supernovae and accelerate relativistic electrons through the magnetic field of the galaxy. Also, in a high-density dusty starburst HII regions around massive stars would be very compact and the ionizing UV absorbed by the high dust column, producing weak nebular emission in the optical. 

To test whether the radio deficiency of NGC~1377 can be explained by a young opaque starburst, we compare some of the observed properties with synthetic starburst models  obtained with the on-line version of {\tt Starburst99}\footnote{\url{http://www.stsci.edu/science/starburst99/}} \citep{starburst99}. We run two sets of models assuming a continuous star formation rate varying from 0.01 to 10~M$_\odot$~yr$^{-1}$ and an instantaneous star formation event with total stellar mass between 10$^4$ and 10$^8$~M$_\odot$. In all models we assume a standard Kroupa initial mass function with exponent 1.3 from 0.1 to 0.5~M$_\odot$~and 2.3 from 0.5 to 100~M$_\odot$. Infrared and mm-wave observations do not show any evidence of a low metallicity for NCG~1377 \citep[e.g., ][]{roussel06} and in our models we consider Z=Z$_\odot$. 

\subsubsection{Non-thermal radio emission}
\label{sec:sb99nt}
The synchrotron emission from supernovae for the {\tt Starburst99} models is shown in Fig.~\ref{fig:sb99}({\it a}). To derive the synchrotron luminosity from the supernova rate SNR we use the conversion L$_\mathrm{NT}$=13$\times$10$^{29}\nu^{-\alpha_\mathrm{NT}}$SNR from \citet{condon92}, with $\nu$ in GHz and $\alpha_\mathrm{NT}$=0.8. As expected, the synchrotron luminosity at 1.4~GHz is zero at times earlier than 3~Myr because no supernova has exploded yet, and in the continuous star formation case it then increases to a roughly constant value at t$>$100~Myr. The shaded area in Fig.~\ref{fig:sb99}({\it a}) shows the confidence interval of the L-band luminosity derived from our JVLA observations for the A component, L$_\mathrm{NT}$=2.8$\pm$0.5$\times$10$^{26}$~erg~s$^{-1}$. We see that the observed value is consistent with either a young starburst with age $<$10~Myr, or with an older system with SFR$\simeq$10$^{-2}$~M$_\odot$~yr$^{-1}$. An instantaneous starburst with total stellar mass between 10$^5$ and 10$^6$~M$_\odot$~and age between 10$^6$ and 10$^7$ could also explain the observed synchrotron flux.

\subsubsection{Ionizing radiation} 
\label{sec:sb99ff}
Free-free emission is produced by HII regions around young massive stars before they explode into supernovae and it is thus a better tracer of the early stages of the starburst compared to synchrotron radiation. Also, free-free radiation is not affected by dust obscuration or, in the limit of optically-thin emission, by the compactness of the HII regions as much as other tracers such as H$\alpha$, which makes it a good tracer of young star formation in obscured environments \citep[e.g., ][]{murphy2011}. 

For optically-thin free-free emission, the thermal radio luminosity is proportional to the production rate $N_\mathrm{UV}$ of Lyman continuum photons from HII regions around young massive stars. Following \citet{condon92}, we can write
\begin{equation}\label{eq:nuv}
N_\mathrm{UV}\geq 6.3\times 10^{52} \left(\frac{T_\mathrm{e}}{10^4~K}\right)^{-0.45}\left(\frac{\nu}{GHz}\right)^{0.1}\left(\frac{L_\mathrm{T}}{W~Hz^{-1}}\right),
\end{equation}
where the inequality accounts for possible absorption of ionizing radiation by dust. Our estimate of the thermal free-free emission for the A component (see Section~\ref{sec:freefree}) results in a thermal radio luminosity at 10~GHz of L$_\mathrm{T}\simeq$6$\times$10$^{18}$~W~Hz$^{-1}$, which when substituted in Eq.~(\ref{eq:nuv}) gives a ionizing photon production rate $N_\mathrm{UV}\simeq$5$\times$10$^{51}$~s$^{-1}$.

In Fig.~\ref{fig:sb99}({\it b}) the $N_\mathrm{UV}$ derived from our observations is compared to the results of {\tt Starburst99} models . The red shaded area shows the observed value, allowing for an uncertainty of 50\%. In the models assuming continuous star formation, $N_\mathrm{UV}$ increases by one order of magnitude from 1 to 10~Myr and is proportional to the SFR. The observed value is consistent either with a young star-formation event with age $<$5~Myr and SFR$\simeq$0.1~M$_\odot$~yr$^{-1}$ or with an older system with SFR of a few 10$^{-2}$~M$_\odot$~yr$^{-1}$. 

For the models assuming instantaneous star formation, $N_\mathrm{UV}$ is roughly constant for the first 2~Myr and then decreases steeply as the massive stars responsible for most of the ionizing UV radiation explode into supernovae. If we assume an age of less than 3~Myr, the observed free-free emission constrains the total stellar mass to $\sim$10$^5$~M$_\odot$. For an older instantaneous starburst the observed value is consistent with ages between a few 10$^6$ and a few 10$^7$~Myr for a total stellar mass range of 10$^6$--10$^8$~M$_\odot$. 

\subsubsection{Bolometric luminosity}

In an opaque starburst, most of the optical and UV stellar radiation is absorbed by dust and re-radiated as thermal continuum in the infrared. In this scenario, the infrared luminosity is a good estimate of the bolometric luminosity of a starburst.

In Fig.~\ref{fig:sb99}({\it c}) we compare the bolometric luminosity of the {\tt Starburst99} models with the observed infrared luminosity of NGC1377. The bolometric luminosity was obtained by integrating the synthetic spectra from 91~$\mbox{\AA}$ to 160~$\mu$m. The total infrared luminosity of NGC~1377 derived from IRAS fluxes \citep{imanishi09} is 1.2$\times$10$^{10}$~L$_\odot$. The observed value is shown in  Fig.~\ref{fig:sb99}({\it c}) as a red shaded area, assuming a confidence interval of 50\%.

For the models assuming continuous star formation, we find that the total IR luminosity is consistent with a young ($<$3~Myr) starburst with SFR$\approx$10~M$_\odot$~yr$^{-1}$, or with a more evolved (>10~Myr) starburst with SFR~$\approx$1~M$_\odot$~yr$^{-1}$. In the case of instantaneous star formation, the bolometric luminosity is dominated by emission by massive stars and declines rapidly after $\sim$3~Myr. For these models, the IR luminosity is consistent with a starburst with mass higher than 10$^7$~M$_\odot$~and age older than 1~Myr, increasing with total stellar mass.

\subsubsection{Mechanical luminosity} \label{sec:mechlum}
Recent ALMA observations of CO~J=3--2 by \citet{aalto2016} reveal a collimated bipolar outflow of molecular gas extending to more than 200~pc above the galactic plane. The de-projected outflow velocity estimates vary between 240 and 850~km/s depending on the assumed inclination, with an estimated upper limit to the H$_2$ mass of 10$^7$~M$_\odot$. This estimate is based on the CO luminosity assuming a standard CO-H$_2$ conversion factor, which may not be applicable to the gas in the outflow. If the gas is turbulent, the conversion factor may be an order of magnitude lower \citep[e.g., ][]{dahmen98}, which would imply an outflow mass as low as 10$^6$~M$_\odot$. Depending on the assumed conversion factor , the mass outflow rate may vary from 1--4~M$_\odot$~yr$^{-1}$~for a {\it light} jet (M(H$_2)=$10$^6$~M$_\odot$) to 10--40~M$_\odot$~yr$^{-1}$~for a {\it massive} jet (M(H$_2)=$10$^7$~M$_\odot$). 

The mechanical luminosity of the outflow can be estimated as 
\begin{equation}
L_\mathrm{mech}=\frac{1}{2}\frac{dM}{dt}V_\mathrm{out}^2, 
\end{equation}
which results in L$_\mathrm{mech}$=2$\times$10$^{40}$--9$\times$10$^{41}$~erg~s$^{-1}$ for the {\it light jet} and 10 times higher for the {\it massive jet}.

In Fig.~\ref{fig:sb99}({\it d}) we compare the mechanical luminosity of the observed molecular outflow with the mechanical luminosity from supernova explosions and stellar winds in the {\tt Starburst99} models. In the case of continuous star formation, we find that in order to produce the observed mechanical luminosity the central starburst should have a SFR of more than 1~M$_\odot$~yr$^{-1}$ and an age of more than 5~Myr.  In particular we find that a pre-supernova starburst cannot drive the observed molecular outflow unless we consider extremely high SFR of a few 10~M$_\odot$~yr$^{-1}$. In the case of instantaneous star formation, we find that the observed mechanical luminosity could be reproduced by a young starburst of total stellar mass $\sim$10$^8$~M$_\odot$~ and age $<$2~Myr. 

In the case of a {\it light jet}, i.e. assuming a CO-H$_2$ conversion factor 10 times lower than the Galactic value, we find that the outflow energetics could be explained by a $>$5~Myr starburst with continuous SFR$>$0.1~M$_\odot$~yr$^{-1}$, or by a young ($<$5~Myr) starburst with SFR $>$1~M$_\odot$~yr$^{-1}$. 

This comparison assumes a perfect coupling between the molecular gas and stellar feedback, i.e. that all the mechanical energy produced by the starburst goes into driving the outflow. Such an assumption is probably not physical and we can expect only a fraction of the mechanical energy from the starburst to be transferred to the molecular outflow. This means that the values derived for the SFR or total stellar mass necessary to drive the outflow should be regarded as lower limits.

\subsection{First X-rays detection ?}
\label{sec:xdet}
Although no X-ray source is detected at the position of the NGC~1377 nucleus (Section \ref{sec:xobs}), we tentatively detect X-ray emission 4$''$.5 to the South-West, with a signal to noise ratio of 2.2 in the 0.3-7~keV band. The position of the X-ray source (\textit{Chandra}'s astrometric accuracy is 0.8$''$) coincides with the position of the radio peak B. 

The signal-to-noise ratio of the X-ray spectrum is insufficient to accurately constrain the spectral shape, but by assuming a spectral model we can deduce the intrinsic X-ray flux. To obtain a flux estimate, an absorbed power law spectrum with a photon index of 1.9, such as is seen for a jet hot-spot \citep{Hardcastle2016} is fitted to the data. The obtained intrinsic luminosity estimates at the 90\,\% confidence limit in the 2-10\,keV band for different columns are given in Table~\ref{col_est}. We find that the intrinsic X-ray luminosity of the B component varies from 10$^{38}$ to 10$^{40}$~erg\,s$^{-1}$ for assumed column densities of 10$^{23}$--10$^{24}$~cm$^{-2}$.

\begin{table}
\caption{\label{col_est} Estimated X-ray luminosities of the B component assuming different hydrogen column densities. See Section \ref{sec:xdet} for discussion.}
\begin{center}
\begin{tabular}{c|c}
\hline
\hline
$N_H$ & $L_{2-10}$ in erg\,s$^{-1}$ \\
$10^{24}\,\mathrm{cm}^{-2}$ & $5.7\times10^{39}-5.1\times10^{40}$\\
$10^{23}\,\mathrm{cm}^{-2}$ & $2.2\times10^{38}-2.0\times10^{39}$ \\
Galactic &  $1.7\times10^{37}-1.5\times10^{38}$ \\
\hline
\hline
\end{tabular}
\end{center}
\end{table}


\section{Discussion}
\label{sec:disc}
\subsection{A nascent starburst ?}

The extremely high IR/radio ratio of NGC~1377 has been interpreted as the signature of a pre-supernova opaque starburst \citep{roussel03,roussel06}. This interpretation was mainly based on the non-detection of the radio continuum by VLA and Effelsberg observations (see Fig.~\ref{fig:sed}). In this scenario the low H$\alpha$ flux would be due to the extreme obscuration by dust, as revealed by deep mid-infrared silicate absorption \citep[e.g., ][]{spoon07}, and by the high pressure which would limit the growth of HII regions. 

\begin{figure}
\includegraphics[width=.9\columnwidth]{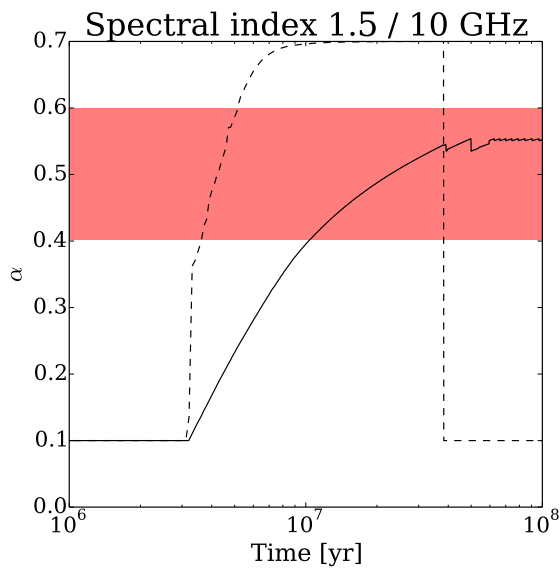}
\caption{\label{fig:spind99} Evolution of the 1.5--10~GHz spectral index from {\tt Starburst99} models (see Sections~\ref{sec:sb99nt} and \ref{sec:sb99ff}). The {\it solid} and {\it dashed} lines represent respectively continuous and instantaneous star formation models. The shaded area shows the confidence interval of the spectral index derived for the NGC~1377 nucleus.}
\end{figure}

Our JVLA observations reveal a steep radio spectral index in the nucleus of NGC~1377, which is usually associated to non-thermal synchrotron emission. In a pre-supernova starburst, the dominant emission mechanism should be thermal free-free from HII regions which would result in a flatter spectrum. This is exemplified by Fig.~\ref{fig:spind99}, where we show the evolution of the radio spectral index with starburst age derived from the {\it Starburst99} models of Section~\ref{sec:sb99}. The observed radio spectral index of 0.5$\pm$0.1 is not consistent with a pre-supernova starburst.

The {\tt Starburst99} models of Section~\ref{sec:sb99} show that  in order to be consistent with the observed IR luminosity, an obscured starburst of age $<$10~Myr should have a SFR higher than 2~M$_\odot$~yr$^{-1}$, or an instantaneous stellar mass greater than 10$^7$~M$_\odot$ (see Fig.~\ref{fig:sb99}({\it c})). 
The synchrotron flux derived by our JVLA observations is consistent with a young starburst mainly because around the threshold for supernova explosion (at about 3~Myr) the models are highly degenerate in SFR (see Fig.~\ref{fig:sb99}({\it a})). A more precise indication of the properties of the putative nascent starburst is given by our estimate of the ionizing photon flux from the observed free-free emission in Fig.~\ref{fig:sb99}({\it b}), which shows that a nascent starburst of age $<$10~Myr should have a SFR$<$0.1~M$_\odot$~yr$^{-1}$, or an instantaneous stellar mass lower that 10$^6$~M$_\odot$~ in order to be consistent with the observations.
The upper limit to the SFR of the nascent starburst is more than one order of magnitude lower than the SFR derived from the IR luminosity. The discrepancy remains in the case of instantaneous star formation, with the radio and IR observations resulting in stellar masses which differ by more than one order of magnitude.

Also, the nascent starburst scenario cannot explain the energetics of the massive molecular outflow observed in CO \citep{aalto1377,aalto2016}. In order to be explained by stellar winds or momentum injection by supernova explosions, the mechanical luminosity of the outflow would require a SFR of 1--10~M$_\odot$~yr$^{-1}$~ or an instantaneous starburst mass of 10$^7$--10$^8$~M$_\odot$, which are one to two orders of magnitude higher than the values estimated from the radio free-free luminosity. Moreover, Starburst-driven outflows are generally wide-angle winds and it is not clear how they could create a highly collimated molecular jet like the one observed in CO 3--2. {\it Therefore, the nascent starburst scenario cannot explain all the observed properties of the central 200~pc of NGC~1377.}

\subsection{An obscured AGN ?}
\label{sec:agn}
\begin{figure}
\begin{center}
\includegraphics[width=.5\textwidth,keepaspectratio]{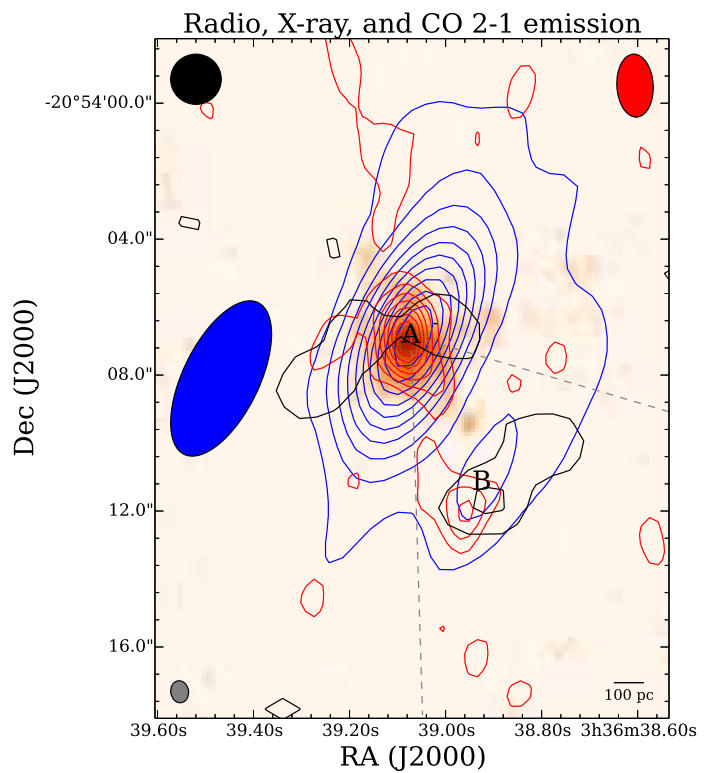}
\caption{\label{fig:comap} Map of radio, CO and X-ray emission in NGC~1377. The radio continuum emission at L ({\it left}) band is shown as {\it blue} contours, the beam is shown in {\it blue}. The radio continuum emission at X ({\it left}) band is shown as {\it red} contours, the beam is shown in {\it red}. The moment~0 map of CO~J=2--1 emission from \citet{aalto1377} is shown in {\it color}, the beam is shown in {\it gray}. The X-rays flux is shown in {\it black} contours, {\it Chandra} smoothed beam is shown in {\it black}. The gray dashed lines show the opening angle of the blue-shifted component of the molecular outflow cone detected in CO by \citet{aalto1377}. }
\end{center}
\end{figure}


The presence of a non-stellar energy source in NGC~1377 was already suggested by \citet{imanishi06} to explain the small observed PAH equivalent width at 3.3~$\mu$m. In addition, the HCN/HCO$^+$ J = 1--0 line ratio exceeds unity, which \citet{imanishi09} interpreted as evidence of an X-ray dominated region (XDR) surrounding an AGN. However, the most convincing evidence came with the detection of the CO molecular outflow by \citep{aalto1377}, who calculated that the upper limit on the 1.4~GHz flux density fell short by a factor of 10 to explain the outflow as supernova-driven and suggested instead that the outflow may be driven by radiation pressure from a buried AGN.
From the calibration of the stellar velocity dispersion and black hole mass \citep[e.g., ][]{graham2011}, \citet{aalto1377} derive a supermassive black hole (SMBH) mass of about 1.5$\times$10$^6$~M$_\odot$. If the IR luminosity of NGC~1377 is produced by accretion onto the SMBH, this should be operating at more than 20\% of its Eddington limit, placing it in the {\it quasar accretion} mode \citep[e.g., ][]{croton2006,janssen2012,mcalpine2015}. 

When the presence of an AGN was first suggested by \citet{imanishi06}, the authors explained the lack of radio emission with a high free-free opacity due to the compact circumnuclear starburst. However, our fit of the radio continuum at L- and X-band finds very low free-free optical depths and a steep spectral index of $\alpha\simeq$0.5-0.6, which are not consistent with a free-free absorbed radio-loud AGN. If an AGN is hiding at the core of NGC~1377 it must be extremely radio faint. The contribution of radio emission to the luminosity of an AGN is often expressed with the radio-loudness parameter $R\equiv L_\mathrm{5 GHz}/L_\mathrm{4400\mbox{\AA}}$. A well established observational result is that optical quasars show a bimodal distribution around $R$=10, with the majority of them being radio-quiet, with 0.1>$R$>1 \citep[e.g., ][]{kellermann1989}. By comparing the optical B band flux density of NGC~1377 \citep[0.012 Jy, ][]{dale2007} with the flux density at 5~GHz (interpolated from our L-band measurement assuming a spectral index $\alpha$=-0.5) we find $R$=0.02, which defines the galaxy as an extremely radio faint source. A well defined anti-correlation between $R$ and the Eddington ratio $\epsilon \equiv L_\mathrm{Bol}/L_\mathrm{Edd}$ is observed \citep[e.g., ][]{sikora2007}. For NGC~1377, the correlation of Fig.~3 by \citet{sikora2007} would predict an Eddington ratio of about unity, which is consistent with the high $\epsilon$ needed to drive the observed outflow and IR luminosity.

We further tested if a 10$^6$\,M$_\odot$ black hole accreting at the Eddington limit is consistent with the non-detection of the nucleus of NGC~1377 in the X-ray band. We performed spectral simulations based on our \textit{Chandra} observation, utilizing the spectrum extracted from the center as a basis, i.e. ensuring that the simulated spectra have the same exposure and same background properties as the actual exposure. The assumed spectral model is an absorbed power law with a photon index of 1.9 with the appropriate normalization for different Eddington ratios ($L/L_\mathrm{edd}$=0.05, 0.1, 0.5, 1.0). To obtain the theoretical 2-10\,keV fluxes for a given Eddington ratio, we used Bolometric corrections from \citet{Vasudevan2007} ($\lambda_{2-10}$=15, 20, 50, 60). The spectra were then simulated using the command \texttt{fakeit} in the X-ray spectral modeling software {\tt xspec} \citep{Arnaud1996}.  We find that a column of $10^{25}$~cm$^{-2}$ is required to bring an AGN accreting at the Eddington limit into agreement with our \textit{Chandra} non-detection (see Table \ref{tab:bhx}). This value is consistent with the lower limit of $6\times10^{24}$~cm$^{-2}$ found with ALMA observations by \citet{aalto2016}. {\it An obscured AGN is thus the most likely explanation for the observed properties of the NGC~1377 nucleus.}

\begin{table}
\begin{center}
\begin{tabular}{cccccc}
\hline 
\hline 
& & & & &\\
& $L/L_\mathrm{edd}$ & 0.05 & 0.1 & 0.5 & 1.0 \\
$N_\mathrm{H}$ & & & & &\\
& & & & &\\
$10^{24}$cm$^{-2}$ &  &  D &   D &  D  &   D \\
$10^{25}$cm$^{-2}$ &  & ND &  ND &  ND &  ND \\
& & & & &\\
\hline 
\hline
\end{tabular}
\end{center}
\caption{\label{tab:bhx} Detectability of a black hole with $M_{BH}=1\times 10^6 M_\odot$ in a \textit{Chandra} ACIS observation with 44\,ks net exposure. The results are shown in dependence of Eddington ratio and column density. D indicates a detection at a signal-to-noise of 2 or greater in the whole \textit{Chandra} band and ND indicates a non-detection.} \label{spec_sim}
\end{table}

\subsection{Is the AGN quenching star formation in NGC~1377?}

In a deep survey of radio sources in the \textit{Chandra} Deep Field South, \citet{padovani2011} found that the evolution of the radio emission from radio-quiet AGN up to a redshift of z$\sim$2.3 is indistinguishable from that of star-forming galaxies, which suggests that most of the emission in radio-quiet AGN must come from star formation. If we assume that all the radio emission from NCG~1377 is powered by star formation, the models of Fig.~\ref{fig:sb99}({\it a}) and ({\it b}) result in an upper limit for the SFR of a few 0.01~M$_\odot$~yr$^{-1}$. This value is more than 20 times lower than the SFR inferred from the SK relation of Section~\ref{sec:sfrind} and could be even lower if we consider that part of the radio emission may be coming from the AGN and not from star formation. {\it The galaxy thus seems to be forming stars at a much lower rate than the local main sequence. }

Feedback from an AGN has been proposed to have a deep effect on the star formation in the host galaxies. In particular, it has been suggested that even low-power AGN could drive massive molecular outflows which could inhibit or shut down entirely star formation in the core of galaxies \citep[e.g., ][]{cicone2014}. 
In a recent study of the low-power AGN host NGC~1266, a system which is very similar to NGC~1377, \citet{alatalo2015} find a star formation rate 50 to 150 times lower than the Schmidt-Kennicutt value. The apparent suppression of star formation may be due to the injection of turbulence by the AGN outflow, which could stabilize the molecular gas against gravitational collapse. 

Following \citet{alatalo2015}, the radial turbulent velocity needed in order to stabilize a spherical molecular bulge is of the order of the free-fall velocity 
\begin{equation}
v_\mathrm{ff}=\sqrt{\frac{3GM_\mathrm{gas}}{5R}},   
\end{equation}
where $M_\mathrm{gas}$ is the gas mass enclosed in the radius $R$. For the inner 60~pc of NGC~1377 we find $v_\mathrm{ff}\approx$50~km\,s$^{-1}$, which is very close to the CO~3-2 line width of 50-60~km\,s$^{-1}$~ measured in the core of NGC~1377 by \citet{aalto2016}. This result suggests that turbulent motions may indeed stabilize the gas against gravitational collapse and hence inhibit star formation in the galaxy nucleus. A velocity dispersion of 50~km\,s$^{-1}$~ could be easily maintained by turbulence generated by the observed AGN outflow \citep[e.g., ][]{appleton2006}, which has a terminal velocity of the order of 240--850~km\,s$^{-1}$. Turbulent energy dissipation would be consistent with the bright mid-IR H$_2$ emission observed by \citet{roussel06}. We note however that there is now significant observational evidence \citep[e.g., ][]{onodera2010,xu2015} that the SK relation may be scale-dependent and may break down at giant molecular cloud (GMC, $\approx$100~pc) scales. The order-of-magnitude difference between the measured SFR and the SK-estimated value in the 60~pc nucleus of NGC~1377 may thus be consequence of the SK breakdown at small scales rather than an indication of negative feedback from an AGN. In order to disentangle the two scenarios, observations at higher angular resolution are needed. These would allow us to assess the contribution of the AGN to the radio flux and derive a more accurate measurement of the SFR surface density.

\subsection{Is the B component associated with NGC~1377?}

A search of publicly available astronomical databases ({\it Simbad}\footnote{\url{http://simbad.u-strasbg.fr/simbad/}}, {\it NED}\footnote{\url{https://ned.ipac.caltech.edu/}}) found no identification of the B radio component as part of an independent Galactic or extragalactic object. The steep radio index makes it unlikely for the B component to be associated with a Galactic source, and optical images do not show any indication of foreground emission. There is however the possibility that the B component may be associated with another radio-faint galaxy.

Following \citet{condonNVSS}, the probability that the nearest unrelated radio source lies within angular distance $r$ of any position is given by
\begin{equation}\label{eq:prob}
P(<r)=1-\exp\left(-\pi\rho r^2\right),
\end{equation}
where $\rho$ is the number density of radio sources. A large-area survey of sub-mJy 1.4~GHz sources \citep[ATESP, ][]{prandoni2000} detected 2960 individual sources brighter than 79~$\mu$Jy in a 26~deg$^2$ field, corresponding to a $\rho_\mathrm{ATESP}\simeq$113~sources/deg$^2$. The B component has an L-band peak flux of 101~$\mu$Jy, which is comparable to the limiting sensitivity of the ATESP survey. If we assume $\rho=\rho_\mathrm{ATESP}$ and $r$=4$''$.5, the probability that the B component is not associated with NGC~1377 but with another extragalactic source is then P(<4$''$.5)=5$\times$10$^{-4}$. 

This probability becomes even lower if we consider the alignment of the B component with the blue-shifted part of the CO outflow detected by \citet{aalto1377}, as shown in Fig.~\ref{fig:comap} and further discussed in the next section.

\subsection{The B component: a hot-spot in the AGN jet?}

The B component shows an optically-thin synchrotron radio spectrum, soft X-ray emission, and is aligned with the blue-shifted part of the molecular outflow detected by \citet{aalto1377,aalto2016} (Fig.~\ref{fig:comap}). If an AGN is indeed powering NGC~1377, the B component may be associated with a hot-spot in an AGN jet. 
Hot-spots are defined as bright compact regions at the end of quasar radio lobes and are usually explained as strong shocks generated by the impact of the relativistic jet with the slow-moving plasma in the lobes \citep[e.g., ][]{blandford74,meisenheimer89}. Both radio and X-ray emission are generated by synchrotron losses as particles are accelerated in the shock. Soft X-ray emission is mostly associated with radio-faint hot-spots, where the magnetic field is low enough to allow particles to be accelerated to X-ray energies \citep[e.g., ][]{Hardcastle2016}. In NGC~1377 we detect only the blue-shifted side of the putative jet, with no detection of a symmetric counter-jet. Asymmetries in the hot-spot positions along a jet are commonly observed \citep[e.g., ][]{mundell2003, Hardcastle2016} which may be due to inhomogeneities in the ISM or magnetic field downstream the jet. The non-detection of a counter-jet hot spot may also indicate that the mean bulk jet speed is still relativistic at the B position, which would enhance the emission of the jet-side hot-spot versus the counter-jet-side one. 

Even if they do not contribute significantly to the bolometric luminosity of the galaxy as in radio-loud sources, relativistic jets are commonly observed in radio-quiet {\it radiative-type} AGNs \citep[e.g., ][]{heckman2014} with high accretion efficiency (such as the putative AGN in NGC~1377, see Section~\ref{sec:agn}) and can have a deep impact on the gas kinematics \citep[e.g., ][]{veilleux2005}. Entrainment of cold molecular gas by a precessing relativistic jet has been suggested by \citet{aalto2016} to explain the extreme collimation of the CO 3-2 outflow in NGC~1377. Hydrodynamical simulations \citep[e.g., ][]{wagner2012} show that 10-40\% of the AGN jet energy can be transferred to the entrained molecular material. The measured mechanical luminosity of the CO outflow ($L_\mathrm{mech}\sim10^{42}$~erg\,s$^{-1}$) is thus a lower limit to the total luminosity of the jet. In radio galaxies, a loose correlation between the total monochromatic power of the AGN at 1.4~GHz  and the luminosity of the relativistic jet is observed \citep[e.g., Eq. 16 in ][]{bizan2008}. By applying this correlation to the measured 1.5 GHz flux of NGC~1377 ($P_\mathrm{1.5GHz}\sim$3$\times$10$^{26}$~erg\,s$^{-1}$/Hz) the expected luminosity of the jet should be of the order of $\sim10^{42}$~erg\,s$^{-1}$, which is similar to the measured mechanical luminosity of the CO outflow ({\it massive jet} case, see Section~\ref{sec:mechlum}). Also, the estimated X-ray luminosity of the B component ({$10^{38}$--$10^{39}$~erg\,s$^{-1}$}, see Section \ref{sec:xdet}) could be easily be powered by such a jet. However, the correlation between AGN power and radio lobe luminosity has a large scatter and recent studies \citep[e.g., ][]{godfrey2016} suggest that this correlation may be very weak when the jet energy is dissipated along its path, e.g. by driving shocks. 

{\it The morphology and energetics of the radio and X-ray emission in NGC~1377 are consistent with an AGN+jet system}. Radio observations at higher resolution will help to confirm the presence of a jet and to study the interactions between the relativistic plasma and the entrained molecular material.

\section{Conclusions}
\label{sec:conc}

We report radio and X-ray observations of NGC1377, the most extreme
FIR-excess galaxy known to date, in which a highly collimated molecular outflow has recently been found. Our results suggest that the morphology and energetics of the radio,
X-ray, and molecular line emissions point toward a radio-faint AGN+jet system explanation, rather than a nascent starburst as previously proposed to interpret the observed properties of the galaxy. Our main results are:
\begin{itemize}
\item We obtained the first detection of the cm-wave radio continuum in NGC~1377. Both the 1.5 and 10~GHz emission show two components, peaking on the galaxy nucleus and 4$''$.5$\sim$500~pc to the South--West. The two radio components have a steep spectral index $\alpha\sim$0.5-0.7, consistent with optically thin synchrotron emission.
\item Soft X-rays emission (0.3-7~keV) is tentatively detected for the first time in NGC~1377 by \textit{Chandra}. The emission is peaked at the position of the off-nucleus radio component with no detection at the galaxy's center. 
\item We confirm the extreme far-IR excess of the galaxy, with a $q_\mathrm{FIR}$ of 4.2 which deviates for more than 7--$\sigma$ from the radio-FIR correlation \citep[$q_\mathrm{FIR}$=2.34$\pm$0.26, ][]{roussel03}.
\item By comparing the observations with synthetic starburst models, we find that the SFR estimated from optically thin free-free ($<$0.1~M$_\odot$~yr$^{-1}$) falls short by 1--2 orders of magnitude from explaining the galaxy's IR luminosity and the mechanical luminosity of the CO outflow. We conclude that a young starburst cannot reproduce all the observed properties of NGC~1377.
\item We estimate that a SMBH of 10$^6$~M$_\odot$ ~accreting at nearly Eddington rates  may reproduce the observed IR luminosity and outflow energy. Such an AGN would be extremely radio-faint, with $R\equiv L_\mathrm{5 GHz}/L_\mathrm{4400\mbox{\AA}}\approx$0.02. 
\item The SFR density measured by radio observations is more than 20 times lower than the SFR inferred from the Schmidt-Kennicutt relation. We find that a turbulent velocity of 50~km/s would be sufficient to stabilize the galaxy bulge against gravitational collapse and inhibit star formation. This value is similar to the velocity dispersion of CO observations. We suggest that turbulent feedback from the AGN may be inhibiting star formation in the bulge of NGC~1377.
\item The radio and X-ray emission from the off-nucleus component are consistent with the presence of a relativistic jet hot-spot. We suggest that this structure may be revealing the presence of a radio counterpart to the CO 3--2 highly collimated outflow. 
\end{itemize}

\begin{acknowledgements}
F.C. acknowledges support from Swedish National Science Council grant 637-2013-7261. A.L. acknowledges support from the ERC Advanced Grant {\it "FEEDBACK"}. R.H.I., M.A.P.T. and A.A. acknowledge support from the Spanish MINECO through grants AYA2012-38491-C02-02 and AYA2015-63939-C2-1-P.
\end{acknowledgements}

\bibliographystyle{aa} 
\bibliography{bibliototale}



\end{document}

%% file: journal.tex
\begin{tabular}{cccccc} 
\hline 
\hline 
& & & & &\\
Frequency & Observation date & Configuration & Beam & Noise r.m.s. & Peak flux density\\
& & & [ maj ($''$) $\times$ min ($''$), PA ($^\circ$) ] & [ $\mu$Jy/beam ] & [ $\mu$Jy/beam ]\\
X band (10 GHz) & October 2014 & C & 4.9 $\times$ 2.2 , -26 & 4 & 82\\
L band (1.4 GHz) & June 2015 & A & 1.8 $\times$ 1.1 , 3 & 10 & 178\\
& & & & &\\
\hline 
\hline
\end{tabular}\\

%% file: source_fit.tex
\begin{tabular}{ccccc} 
\hline 
\hline 
& & & & \\
Component & Position & Deconvolved size & Peak flux density & Flux density \\
& [ R.A. (h:m:s) , Dec. ($^\circ:':''$) ] & [ maj ($''$) $\times$ min ($''$), PA ($^\circ$) ] & [ $\mu$Jy/beam ] & [ $\mu$Jy ] \\
& & & &\\
X-band A & 3:36:39.06 , -20:54:06.8 & 4.4$\pm$0.7 $\times$ 2.5$\pm$0.3, 163$\pm$12 & 80$\pm$4 & 160$\pm$12 \\
X-band B & 3:36:38.93 , -20:54:10.9 & point source & 28$\pm$4 & 25$\pm$2\\
& & & &\\
L-band A & 3:36:39.07 , -20:54:06.8 & 2.6$\pm$0.7 $\times$ 1.7$\pm$0.6, 26$\pm$68 & 179$\pm$21 & 427$\pm$69\\
L-Band B & 3:36:38.96 , -20:54:11.8 & point source & 101$\pm$21  & 101$\pm$19\\
& & & &\\
\hline 
\hline
\end{tabular}\\

%% file: sfr_table.tex
\begin{center}
\begin{tabular}{ll|lr} 
\hline 
\hline 
& & &\\
L$_\mathrm{NT}$(1.5~GHz)$^{(a)}$ & 2.8$\times$10$^{26}$~erg~s$^{-1}$~Hz$^{-1}$ & $^{(f)}$SFR$_\mathrm{NT}$=6.64$\times 10^{-29} \nu^{\alpha_\mathrm{NT}}$ L$_\mathrm{NT}$ =  & 0.02 M$_\odot$~yr$^{-1}$\\ 
L$_\mathrm{T}$(1.5~GHz)$^{(b)}$ & 7.1$\times$10$^{25}$~erg~s$^{-1}$~Hz$^{-1}$ & $^{(g)}$SFR$_\mathrm{T}$=4.6$\times 10^{-28}\left(T_\mathrm{e}/10^4~K\right)^{-0.45}\nu^{0.1}$ L$_\mathrm{T}$= & 0.03 M$_\odot$~yr$^{-1}$\\ 
L(24~$\mu$m)$^{(c)}$ & 1.2$\times$10$^{30}$~erg~s$^{-1}$~Hz$^{-1}$ & $^{(h)}$SFR$_{24\mu\mathrm{m}}$=5.58$\times 10^{-36}(\nu $L(24~$\mu$m)$)^{0.826}$=  & 2.6~M$_\odot$~yr$^{-1}$\\
L$_\mathrm{IR}$$^{(d)}$ & 4.8$\times$10$^{43}$~erg~s$^{-1}$ & $^{(i)}$SFR$_\mathrm{IR}$=3.88$\times$10$^{-44}$L$_\mathrm{IR}$= & 1.9~M$_\odot$~yr$^{-1}$\\
$\Sigma_\mathrm{gas}$$^{(e)}$ & 1.4$\times$10$^4$~M$_\odot$~pc$^{-2}$ & $^{(k)}$SFR$_\mathrm{SK}$=$2.5\times10^{-4}\Sigma_\mathrm{gas}^{1.4}\times\pi\left(d/2\right)^2$= & 0.45~M$_\odot$~yr$^{-1}$\\
& & &\\
\hline 
\hline
\end{tabular}\\
\end{center}
{\tiny\it a) Radio luminosity at 1.5 GHz. b) Thermal radio luminosity at 1.5~GHz from the fit of Section~\ref{sec:freefree}. c) Luminosity at 24~$\mu$m, from \citet{temi2009}. d) Total infrared luminosity from IRAS fluxes \citep{imanishi09}. e) Molecular gas surface density from \citep{aalto1377} assuming a size of 200~pc. f) SFR from non-thermal radio emission, Eq. 14 in \citet{murphy2011}. g) SFR from thermal radio emission, Eq. 11 in \citet{murphy2011}. h) SFR from  24~$\mu$m emission,  Eq. 5 in \citet{murphy2011}. i) SFR estimate from IR luminosity, Eq. 4 in \citet{murphy2011}. k) SFR from molecular gas surface density, assuming a Schmidt-Kennicutt law with exponent 1.4 and a disk diameter $d$=60~kpc. }